\documentclass[aps,prev,twocolumn,preprintnumbers,floatfix,nofootinbib]{revtex4-2}
\pdfoutput=1
\usepackage[english]{babel}
\usepackage{graphicx,color}
\usepackage{bm}
\usepackage{times}
\usepackage{slashed}
\usepackage{multirow}
\usepackage[usenames,dvipsnames,svgnames,table]{xcolor}
\usepackage{slashed}
\usepackage{amsmath}
\usepackage{enumitem}
\usepackage{amsthm}
\usepackage{subfigure}
\usepackage{hyperref}
\definecolor{bluencs}{rgb}{0.0, 0.53, 0.74}
\definecolor{darkcyan}{rgb}{0.0, 0.55, 0.55}
\definecolor{hanblue}{rgb}{0.27, 0.42, 0.81}
\definecolor{blue2}{RGB}{53, 56, 170}
\usepackage{makecell}
\usepackage{amssymb}
\newcommand{\orcid}[1]{\hspace{1mm}\href{https://orcid.org/#1}{\includegraphics[height=0.3cm,keepaspectratio]{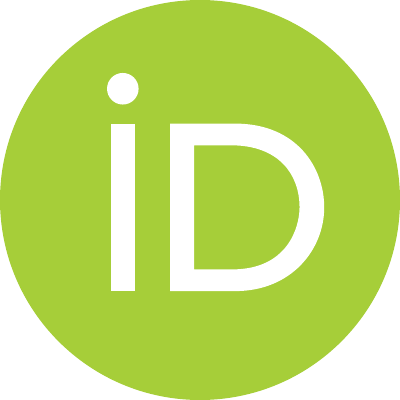}}}
\usepackage{xcolor} 
\usepackage[normalem]{ulem}

\definecolor{green}{rgb}{0.0, 0.5, 0.0}
\usepackage{booktabs}
\usepackage{makecell}      
\usepackage{siunitx}        
\usepackage{adjustbox}       
\usepackage{graphicx}
\hypersetup{
	linktocpage=true,
	setpagesize=true,
	urlcolor=darkcyan,
	citecolor=darkcyan,
	linkcolor=darkcyan, 
	menucolor=cyan,
	colorlinks=true, 
	filecolor=darkcyan,
	citebordercolor=darkcyan,
	pdftitle={Interpreting the 650 GeV and 95 GeV Higgs Anomalies in the  N2HDM},
	pdfsubject={Latex},
	pdfauthor={Benbrik, Boukidi, Moretti, Kahime, Rahili, Taki},
	pdfkeywords={BSM},
	unicode=true
}
\usepackage[capitalize]{cleveref}

\newcommand{\be}{\begin{equation}}
	\newcommand{\ee}{\end{equation}}
\newcommand{\bea}{\begin{eqnarray}}
	\newcommand{\eea}{\end{eqnarray}}

\newcommand{\beq}{\begin{eqnarray}}
	\newcommand{\eeq}{\end{eqnarray}}
\newcommand{\bpmatrix}{\begin{pmatrix}}
	\newcommand{\epmatrix}{\end{pmatrix}}
\newcommand{\ba}{\begin{array}}
	\newcommand{\ea}{\end{array}}

\usepackage{xfrac}
 
\usepackage{float}
\usepackage{bbding,pifont}
\definecolor{brilliantrose}{rgb}{1.0, 0.33, 0.64}
\definecolor{lawngreen}{rgb}{0.49, 0.99, 0.0}
\definecolor{magenta}{rgb}{1.0, 0.0, 1.0}
\makeatletter
\let\old@float\@float
\def\@float{\let\centering\relax\old@float}
\makeatother
\begin{document}
	\preprint{IFJPAN-IV-2026-6}
	\title{Interpreting the 650 GeV and 95 GeV Higgs anomalies \\ in the next-to-two-Higgs-doublet model}

	\author{R. Benbrik$^{1}$\orcid{0000-0002-5159-0325}}
	\email{r.benbrik@uca.ac.ma}	
	\author{M. Boukidi$^{2}$\orcid{0000-0001-9961-8772}}
	\email{mohammed.boukidi@ifj.edu.pl}
	\author{K. Kahime\orcid{0000-0001-8013-3521}$^{3}$}
	\email{Kahimek@gmail.com}
	\author{S. Moretti$^{4,5}$\orcid{0000-0002-8601-7246}}
	\email{s.moretti@soton.ac.uk}\email{stefano.moretti@physics.uu.se}
	\author{L. Rahili$^{6}$\orcid{0000-0002-1164-1095}}
	\email{rahililarbi@gmail.com}
	\author{B. Taki$^{6}$\orcid{0009-0009-2642-1288}}
	\email{taki.bassim@edu.uiz.ac.ma}

	\affiliation{$^1$Polydisciplinary Faculty, Laboratory of Fundamental and Applied Physics, Cadi Ayyad University, Sidi Bouzid, B.P. 4162, Safi, Morocco.\\	$^2$Institute of Nuclear Physics, Polish Academy of Sciences, ul. Radzikowskiego 152, Cracow, 31-342, Poland.\\	$^3$Laboratoire Interdisciplinaire de Recherche en Environnement, Management, Energie et Tourisme (LIREMET), ESTE, Cadi Ayyad University, B.P. 383, Essaouira, Morocco.\\	$^4$School of Physics and Astronomy, University of Southampton,\\ Southampton, SO17 1BJ, United Kingdom.\\$^5$Department of Physics and Astronomy, Uppsala University, Box 516, SE-751 20 Uppsala, Sweden.\\$^6$Laboratory of Theoretical and High Energy Physics (LPTHE), Faculty of Sciences, Ibnou Zohr University, B.P 8106, Agadir, Morocco.}
		
\begin{abstract}
Recent experimental hints from the Large Hadron Collider (LHC) in di-photon  and partially in the $\tau^+\tau^-$ final states suggest the possible existence of an additional Higgs boson with a mass around 95 GeV. Interestingly, these observations are consistent with earlier results from the Large Electron-Positron (LEP) collider, which  pointed to an excess in $b\bar b$ final states within a similar mass range. Additionally, CMS has observed an excess in the  $\gamma\gamma b\bar{b}$ final state, indicating a possible resonance near 650 GeV decaying into a pair of SM-like Higgs bosons or into a SM-like Higgs boson accompanied by a lighter scalar with mass near 95 GeV. In this work, we investigate whether these anomalies can be simultaneously explained within the Next-to-2-Higgs-Doublet Model (N2HDM), an extension of the Standard Model (SM) scalar sector featuring two complex Higgs doublets and an additional real singlet. Assuming the existence of a CP-even Higgs state compatible with the 95 GeV excesses (restricted to the $\gamma\gamma$ and $b\bar b$ channels), we analyse the Type-II and Type-Y Yukawa structures, taking the observed 650 GeV resonance to be a CP-even Higgs state. An extensive parameter scan is performed, incorporating the latest constraints from the properties of the observed 125 GeV Higgs boson, direct searches for additional Higgs states, flavour physics data, and Electroweak Precision Observables (EWPOs). Our results show that a heavy CP-even Higgs resonance around 650 GeV, produced predominantly via gluon-gluon fusion and subsequently decaying into a 125 GeV Higgs boson together with another scalar at approximately 95 GeV, can be simultaneously accommodated within both the N2HDM Type-II and Type-Y frameworks in parameter regions that remain consistent with the relevant experimental $2\sigma$ intervals for the reported excesses, once all theoretical and experimental constraints are imposed. This interpretation leads to distinctive and testable predictions for the ongoing LHC Run~3 and the forthcoming High-Luminosity LHC (HL-LHC) phase, in particular through correlated rates in the $\gamma\gamma b\bar b$, $\tau^+\tau^- b\bar b$, $b\bar b\,\gamma\gamma$, and $\gamma\gamma\tau^+\tau^-$ final states.
\end{abstract}
		
		\maketitle
		\section{Introduction}
		\label{sec:intro}
		
		The discovery of a 125 GeV Higgs boson at the LHC in July 2012 ($H_{\rm SM}$)~\cite{CMS:2012qbp, ATLAS:2012yve} confirmed the key missing piece of the Standard Model (SM) and, most crucially, confirmed the strong role of the Brout-Englert-Higgs mechanism of Electro-Weak Symmetry Breaking (EWSB). However, despite this success, the SM Higgs sector faces several theoretical shortcomings that motivate extensions. In particular, the SM provides no explanation for key open questions such as neutrino masses, dark matter, and the matter-antimatter asymmetry from the experimental side as well as the naturalness (or hierarchy) problem on the theoretical side, among others. These limitations have spurred considerable interest in models with an extended Higgs sector, viewed as low-energy realizations of more fundamental high-scale theories. 
		
		Extending the scalar sector of the SM can alleviate many of the issues mentioned above. The presence of additional scalar fields can introduce new interactions that improve the high-scale behavior of the Higgs potential or provide hints of new physics Beyond the SM (BSM), potentially addressing the naturalness problem altogether. While non-minimal supersymmetric models---e.g., the Next-to-Minimal Supersymmetric Standard Model (NMSSM)~\cite{Ellwanger:2009dp} or the new Minimally-extended Supersymmetric Standard Model (nMSSM)~\cite{Dedes:2000jp} (see~\cite{Moretti:2019ulc} for a review of a class of these)---are compelling solutions to this issue, as they typically introduce a large array of new particles and parameters. An alternative, more minimal and non-supersymmetric approach is to extend the SM by a second $SU(2)_L$ scalar doublet, known as the 2-Higgs-Doublet Model (2HDM)~\cite{Gunion:1992hs,Branco:2011iw}, or by adding both a Higgs doublet and a real singlet field. The latter is known as the N2HDM~\cite{Muhlleitner:2016mzt,Chen:2013jvg,Muhlleitner:2017dkd,Ferreira:2019iqb,Engeln:2020fld,Arhrib:2018qmw,Arhrib:2024itt,Binjonaid:2024gwo}, providing a simpler alternative to the non-minimal supersymmetric scenarios. In fact, it captures many of the same Higgs-sector benefits (additional states, a singlet that can help explain observed resonances, etc.) without the elaborate superpartner sector. This simplicity implies fewer parameters and often clearer correlations among observables, making the N2HDM an appealing framework for interpreting collider anomalies. 
		
		In fact, recent experimental observations at the LHC and previous colliders have revealed intriguing hints of new (pseudo)scalar resonances beyond the well-established 125 GeV state. Among the most notable ones are the excesses around 95 GeV, first seen in LEP experiments and later reinforced by the CMS and ATLAS collaborations in multiple decay channels including $b\bar{b}$~\cite{ALEPH:2006tnd}, 
		$\gamma\gamma$~\cite{CMS:2018cyk,CMS:2024yhz,ATLAS:2024bjr,ATLAS:2023jzc} plus  $\tau^+\tau^-$~\cite{CMS:2022goy}, which have received considerable attention recently and have  been extensively studied in various theoretical frameworks~\cite{Cao:2016uwt,Heinemeyer:2021msz,Biekotter:2021qbc,Biekotter:2019kde,Cao:2019ofo,Biekotter:2022abc,Iguro:2022dok,Li:2022etb,Cline:2019okt,Biekotter:2021ovi,Crivellin:2017upt,Cacciapaglia:2016tlr,Abdelalim:2020xfk,Biekotter:2022jyr,Biekotter:2023jld,Azevedo:2023zkg,Biekotter:2023oen,Cao:2024axg,Wang:2024bkg,Li:2023kbf,Dev:2023kzu,Borah:2023hqw,Cao:2023gkc,Aguilar-Saavedra:2023tql,Ashanujjaman:2023etj,Dutta:2023cig,Ellwanger:2024txc,Diaz:2024yfu,Ellwanger:2024vvs,Ayazi:2024fmn,Coloretti:2023wng,Bhattacharya:2023lmu,Ahriche:2023hho,Ahriche:2023wkj,Benbrik:2022azi,Benbrik:2022tlg,Belyaev:2023xnv,Janot:2024lep,Gao:2024ljl,Benbrik:2024ptw,Li:2025tkm,Hmissou:2025uep,Gao:2024qag,Dutta:2025nmy,Abbas:2025ser,Xu:2025vmy,Arhrib:2025pxy,Coutinho:2024zyp,Abbas:2024jut,Baek:2024cco,Banik:2024ugs,Mondal:2024obd,Dong:2024ipo,Robens:2024wbw,BrahimAit-Ouazghour:2024img,Khanna:2024bah,Janot:2024ryq,YaserAyazi:2024hpj,Ogreid:2024qfw,Du:2025eop,Lian:2024smg,Chang:2025bjt,Kundu:2024sip,Mondal:2025tzi}. Another prominent anomaly has been observed by the CMS collaboration near 650 GeV in the di-photon plus $b\bar{b}$ final state~\cite{CMS:2023boe}, suggesting a heavy resonance potentially decaying into a pair of Higgs bosons or a Higgs boson accompanied by a $Z$ boson. Indeed, in the absence of any observed excess in the $\tau^+\tau^- b\bar{b}$ final state, the CMS collaboration has placed a 95$\%$ Confidence Level (CL) upper limit of approximately 3 $\rm fb$ on the cross section for a 650 GeV resonance decaying via cascade processes such as $X_{650} \to  H_{\rm SM}  Y_{95} \to \tau^+\tau^- b\bar{b}$~\cite{CMS:2021yci}. This constraint, derived from full Run 2 data at 13 TeV, serves as a critical test when evaluating scenarios involving heavy Higgs cascades. Hence, any viable model interpretation must ensure that the predicted cross section for this channel remains below this threshold in order to be consistent with all current experimental bounds. In addition to these two anomalies, further mild excesses associated with charged Higgs states have also been investigated~\cite{Akeroyd:2022ouy,Arhrib:2024sfg,Bernal:2023aai}, providing additional motivation to investigate extended Higgs sectors. 
		In the literature, many theoretical frameworks have been proposed to simultaneously accommodate the 650 GeV and 95 GeV resonances, including the N2HDM with a $U(1)_H$ gauge symmetry~\cite{Banik:2023ecr}, the NMSSM~\cite{Ellwanger:2023zjc}, the 2HDM Type-III~\cite{Benbrik:2025hol}, the 2HDM Type-I~\cite{Khanna:2025cwq}, and the I(1+2)HDM~\cite{Hmissou:2025riw}. 
		
		Motivated by these considerations and benefiting from the mixing of the additional singlet scalar with the two doublets, the N2HDM allows for more flexibility in accommodating multiple (pseudo)scalar resonances, thereby providing a consistent and simultaneous interpretation of both the 95 GeV and 650 GeV anomalies.
		In this study, we systematically analyze the parameter space of the N2HDM within the Type-II and Type-Y Yukawa structures, considering two distinct scenarios in which the 650 GeV resonance corresponds to either a CP-even (scalar) or a CP-odd (pseudoscalar) state. In the first scenario, a particularly compelling interpretation is that the heavier CP-even state ($h_3$) near 650 GeV decays directly into the SM-like Higgs boson ($h_2\equiv H_{\rm SM}$) and a lighter Higgs boson ($h_1$) around 95 GeV, followed by $h_2$ decays to $\gamma\gamma$ and $h_1$ decays to $b\bar{b}$. Alternatively, the 650 GeV resonance may correspond to the CP-odd Higgs state ($A$) of the N2HDM, which allows for a cascade decay into a SM-like Higgs boson ($h_2$) and a $Z$ boson, yielding the same final state with two photons and two bottom (anti)quarks\footnote{Note that, although the cited  CMS analysis of the $\gamma\gamma b\bar b$ signal  models the excess as a decay $X_{650} \to  H_{\rm SM}  Y_{95}$, with $X_{650}$ and $Y_{95}$ being spin-0 states, 
			at $\sqrt s \approx m_A = 650$ GeV,
			where $m^2_{Z}/s\ll 1$, the equivalence theorem affirms that the pseudoscalar polarization of the $Z$ boson behaves like the
			corresponding neutral Goldstone mode, which is indeed a (CP-odd) spin-0 state, 
			with deviations suppressed by ${\cal O}(m^2_Z /s)\leq 2\%.$}, respectively. Furthermore, recall that, given the limited resolution of the $b\bar b$ and $\tau^+\tau^-$ invariant masses (of order 10 GeV) with respect to the $\gamma\gamma$ one (of order 1 GeV) at the above LEP and LHC experiments, all such excesses could well  be consistent with the best measured value (in the di-photon channel) of 95 GeV (so that, in this paper, we assume the latter value as the common one to all anomalies).   
		
		By incorporating all relevant experimental constraints, we find that the CP-even interpretation of the 650~GeV resonance remains viable in both N2HDM Type-II and Type-Y frameworks, with the corresponding predictions lying within the adopted $2\sigma$ experimental intervals for the 95 GeV and 650 GeV hints. 
		In contrast, the CP-odd interpretation yields no parameter region that simultaneously accounts for both excesses
		once the existing LHC searches for heavy neutral Higgs bosons are imposed mainly by direct $A \to h_2 Z$ searches~\cite{CMS:2019ogx, ATLAS:2022enb, ATLAS:2020gxx}, together with $t\bar{t}Z-$sensitive top-associated modes~\cite{CMS:2024yiy}, and subleading constraints from the $\tau^+\tau^-$ searches~\cite{ATLAS:2020zms}. As a result, the pseudoscalar option is effectively excluded by current BSM Higgs limits.
		Accordingly, the remainder of this work focuses exclusively on the CP-even scenario, for which the lightest CP-even state near 95 GeV remains consistent with the aforementioned $\gamma\gamma$ and $b\bar{b}$ excesses.
		These findings offer valuable guidance for future searches, indicating that the 650 and 95 GeV excesses may be simultaneously explained within extended scalar sectors, pointing toward BSM physics (as hinted by the extensive literature on the subject).
		
		The paper is organized as follows. Section~\ref{sec:n2hdm_nutshell} briefly reviews the main features of the N2HDM setup. In Section~\ref{sec:constraints}, we outline the theoretical and experimental constraints used to shape the parameter space of this scenario. Section~\ref{sec:anomalies} summarizes the relevant experimental anomalies. In Section~\ref{sec:results}, we present the numerical results for the CP-even scenario and highlight selected Benchmark Points (BPs). We conclude in Section~\ref{sec:conlusion}.
		
		\section{Basics of the N2HDM}
		\label{sec:n2hdm_nutshell}
		
		In this section, we briefly discuss the basic features of the N2HDM, which extends the SM Higgs sector by incorporating two complex $SU(2)_L$ Higgs doublets $H_{i}$ ($i = 1,2$), with hypercharge $Y=1$, and a real Higgs singlet, with hypercharge $Y=0$. This extension increases the (pseudo)scalar field content, introducing additional physical Higgs bosons and expanding the phenomenological scope of the model. After EWSB, the fields acquire Vacuum Expectation Values (VEVs) and can be parametrized as
		\begin{eqnarray}
			H_{i} & = \left(
			\begin{array}{c}
				\phi_i^\pm \\
				\frac{1}{\sqrt{2}}(v_i + \phi_i + i \chi_i) \\
			\end{array}
			\right)~~{\rm and}~~S = v_s + \phi_s,
		\end{eqnarray}
		where $v_1$, $v_2$ and $v_s$ denote the VEVs of the respective fields, satisfying $v=\sqrt{v_1^2+v_2^2}=246$ GeV.
		
		The N2HDM Higgs potential, consistent with the $SU(2)_L\otimes U(1)_Y$ gauge symmetry, is given by~\cite{Muhlleitner:2016mzt,Chen:2013jvg,Muhlleitner:2017dkd,Ferreira:2019iqb,Engeln:2020fld,Arhrib:2018qmw,Arhrib:2024itt,Binjonaid:2024gwo}:
		%
        %
        {\small \begin{align}
        	V(H_1,H_2,S) &= m_{11}^2\, H_1^\dagger H_1
        	+ m_{22}^2\, H_2^\dagger H_2
        	- \mu_{12}^2\!\left(H_1^\dagger H_2 + H_2^\dagger H_1\right)
        	 \nonumber\\
        	&\quad + \frac{1}{2}m_S^2 S^2 + \frac{\lambda_1}{2}\left( H_1^\dagger H_1 \right)^2 
        	+ \frac{\lambda_2}{2}\left( H_2^\dagger H_2 \right)^2
        	 \nonumber\\
        	&\quad + \lambda_3\, (H_1^\dagger H_1)(H_2^\dagger H_2) + \lambda_4\, (H_1^\dagger H_2)(H_2^\dagger H_1)
        	 \nonumber\\
        	&\quad + \frac{\lambda_5}{2}\Big[\left( H_1^\dagger H_2 \right)^2
        	+ \left( H_2^\dagger H_1 \right)^2 \Big] \nonumber\\
        	&\quad + \frac{\lambda_6}{8} S^4
        	+ \frac{1}{2}\Big(\lambda_7 H_1^\dagger H_1
        	+ \lambda_8 H_2^\dagger H_2\Big) S^2,
        	\label{eq:Vpot}
        \end{align}
        } 
		where $m_{11}^2$, $m_{22}^2$ and $m_{S}^2$ are the quadratic mass terms for the Higgs doublets $H_{1,2}$ and the singlet field $S$, respectively. The CP conservation of the scalar potential, which we assume, is guaranteed by the absence of complex phases in the parameters (including the VEVs). Accordingly, all quartic couplings $\lambda_i$ ($i=1,\dots,8$) and the bilinear term $\mu_{12}^2$ are assumed to be real.
		
		This potential structure arises from the imposition of two discrete $\mathbb{Z}_2$ symmetries. In general, if both doublets couple to the same fermion types, tree-level Flavor Changing Neutral Currents (FCNCs) arise, which are strongly constrained by experiment. To eliminate such effects, a discrete $\mathbb{Z}_2$ symmetry is imposed under which the fields transform as $H_1 \to H_1$, $H_2 \to -H_2$ and $S \to S$. This symmetry is softly broken by the $\mu_{12}^2$ term to avoid undesirable vacua and ensure a realistic Higgs spectrum.
		A second $\mathbb{Z}_2$ symmetry, under which $S \to -S$, is also required to distinguish the singlet role. This symmetry is spontaneously broken once $S$ develops a VEV, in turn  allowing singlet-doublet mixing and thus enriching the scalar phenomenology compared to the 2HDM, while essentially leaving untouched the pseudoscalar and charged Higgs sector of the latter.
		
		The physical Higgs spectrum of the N2HDM consists of three CP-even neutral scalars ($h_1$, $h_2$, and $h_3$ with $m_{h_1} < m_{h_2} < m_{h_3}$), one CP-odd neutral scalar ($A$) and a charged Higgs pair ($H^\pm$). The squared masses of the charged and CP-odd Higgs states match those in the 2HDM and are related via~\cite{Gunion:2002zf}:
		\begin{eqnarray}
			m_{H^\pm}^2 = m_{A}^2 + \frac{1}{2} v^2 (\lambda_5 - \lambda_4),
			\label{eq:mAmHp}
		\end{eqnarray}
		with the associated mass matrices diagonalized by the rotation
		\begin{equation}
			\label{eq:rotation-matx}
			{{R}}_{\beta}=\begin{pmatrix}
				c_\beta & s_\beta  \\
				-s_\beta & c_\beta
			\end{pmatrix},
		\end{equation}
		where $c_\beta \equiv \cos\beta$, $s_\beta \equiv \sin\beta$ and $\tan\beta \equiv t_\beta = v_2/v_1$.
		
		In contrast, the CP-even sector differs from that of the 2HDM due to the aforementioned mixing. The three physical CP-even scalars $h_i = \{h_1,\ h_2,\ h_3\}$ arise from the orthogonal rotation:
		\begin{eqnarray}
			\left(
			\begin{array}{c}
				h_1\\
				h_2\\
				h_3
			\end{array}
			\right) = {{R}}\,\left(
			\begin{array}{c}
				\phi_1\\
				\phi_2\\
				\phi_s
			\end{array}
			\right),
		\end{eqnarray}
		where ${{R}}$ is a $3 \times 3$ rotation matrix parametrized by the three mixing angles $\alpha_1$, $\alpha_2$ and $\alpha_3$:
		\begin{equation}
			\label{eq:RotMat}
			{{R}} = 
			\left(
			\begin{array}{ccc}
				c_1 c_2 & s_1 c_2  & s_2 \\
				-c_1 s_2 s_3 - s_1 c_3 & c_1 c_3 - s_1 s_2 s_3 & c_2 s_3 \\
				-c_1 s_2 c_3 + s_1 s_3  & -c_1 s_3 - s_1 s_2 c_3 & c_2 c_3
			\end{array}
			\right),
		\end{equation}
		with $s_k = \sin(\alpha_k)$ and $c_k = \cos(\alpha_k)$ for $k = 1,2,3$.
		
		This mixing leads to a non-trivial singlet-doublet composition of the CP-even Higgs states, quantified by $\Sigma_{h_i} = |R_{i3}|^2$ ($i=1,2,3$), which denotes the singlet fraction in each mass eigenstate. This  admixture modifies the couplings of the Higgs bosons to SM particles, thereby affecting their Branching Ratios (BRs) and production cross sections.
		In the context of the present analysis, this singlet-doublet mixing plays a central role: it allows $h_1$ to have sufficiently suppressed couplings to electroweak gauge bosons to be compatible with the LEP $Zh_1$ rate, while still retaining adequate couplings to fermions and gluons to reproduce the LHC $\gamma\gamma$ and $b\bar b$ signals, and at the same time permits $h_3$ to develop a sizeable $h_3\to h_2 h_1$ branching ratio without spoiling the SM-like nature of $h_2$. As we show in Sec.~\ref{sec:results}, the parameter points that survive all constraints share common trends: moderate $\tan\beta$, a suppressed heavy-to-$VV$ coupling, a dominant $h_1\to b\bar b$ decay and a non-negligible singlet fraction in both $h_1$ and $h_3$.
		
		Regarding Yukawa interactions, the first $\mathbb{Z}_2$ symmetry is extended to the fermion sector to ensure that each type of fermion couples to only one Higgs doublet, thereby forbidding FCNCs at tree level. This leads to four distinct Yukawa structures, as follows.
		\begin{enumerate}[label=(\roman*)]
			\item Type-I: All fermions couple to $H_2$.
			
			\item Type-II: Up-type quarks couple to $H_2$, down-type
			quarks and leptons to $H_1$.
			
			\item Type-X (Lepton-specific): Quarks couple to $H_2$,
			leptons to $H_1$.
			
			\item Type-Y (Flipped): Up-type quarks and leptons
			couple to $H_2$, down-type quarks to $H_1$.
		\end{enumerate}
		
		The couplings of the CP-even Higgs states to fermions, normalized to their SM values can be expressed in terms of the rotation matrix elements $R_{ij}$, as shown in Table~\ref{tab:yukcoup}. Notably, the Yukawa structure remains identical to that of the 2HDM \cite{Branco:2011iw}. 

		\begin{table}[htpb!]
			\centering
			\renewcommand{\arraystretch}{1.0}  
			\setlength{\tabcolsep}{16pt} 
			\begin{adjustbox}{max width=0.5\textwidth}
				\begin{tabular}{lccc} \Xhline{0.85pt}
					& $c_{h_i t\bar t}$ \qquad& $c_{h_i b\bar b}$ \qquad&
					$c_{h_i \tau\bar\tau}$ \\ \hline
					Type~I \qquad & ${{R}}_{i2} / s_\beta$
					\qquad& ${{R}}_{i2} / s_\beta$ \qquad&
					${{R}}_{i2} / s_\beta$ \\
					Type~II \qquad & ${{R}}_{i2} / s_\beta $
					& ${{R}}_{i1} / c_\beta $ &
					${{R}}_{i1} / c_\beta $ \\
					Type~X \qquad & ${{R}}_{i2} / s_\beta$
					& ${{R}}_{i2} / s_\beta$ &
					${{R}}_{i1} / c_\beta$ \\
					Type~Y \qquad & ${{R}}_{i2} / s_\beta$
					& ${{R}}_{i1} / c_\beta$ &
					${{R}}_{i2} / s_\beta$ \\ \Xhline{0.85pt}
				\end{tabular}
			\end{adjustbox}
			\caption{The normalized (to the SM) Yukawa couplings of the N2HDM Higgs bosons $h_i$ ($i=1,2,3$).}
			\label{tab:yukcoup}
		\end{table}
		
		Moreover, by expanding the covariant derivatives in the kinetic term of the Lagrangian, one can easily derive the coupling coefficients
		for the couplings to the gauge bosons $W^{\pm}$ and $Z$, for the three CP-even Higgs bosons. These reduced couplings are independent of the Yukawa type and can be expressed as: 
		\begin{equation}
			c_{h_i VV} = c_\beta {{R}}_{i1} + s_\beta {{R}}_{i2},
			\label{eq:n2hdmgaugecoup}
		\end{equation}
		which satisfy the following sum rule:
		\begin{eqnarray}
			\sum_{i=1}^3 c_{h_i VV}^2 =1.
		\end{eqnarray}
		We further notice here that the singlet component ${{R}}_{i3}$ does not couple directly to gauge bosons since the singlet does not contribute to the gauge interaction terms in the kinetic part of the Lagrangian. 
		
		For our analysis, we use the public code {\tt ScannerS} \cite{Coimbra:2013qq,Muhlleitner:2020wwk,Muhlleitner:2016mzt} to set the N2HDM inputs in terms of the reduced couplings and mixing matrix elements, adopting its scheme with the SM-like state $H_{\rm SM}$  identified as $ h_2$. The independent input parameters are:
		\begin{equation}
			\label{eq:inputs}
			\begin{aligned}[t]
				& m_{h_{1}},\ m_{h_{2}},\ m_{h_{3}},\ m_A,\ m_{H^\pm},\ \tan\beta, \\
				& c_{h_2VV}^2,\ c_{h_2 t\bar t}^2,\ \mathrm{sign}(R_{23}),\ R_{13},\ m_{12}^2,\ v_s .
			\end{aligned}
		\end{equation}
		Here $c_{h_2 VV}^2$ and $c_{h_2 t\bar{t}}^2$ are the squared reduced couplings of $h_2$ fixed by the assumption $(c_{h_2 VV}^2)(c_{h_2 t\bar{t}}^2) > 0$ to determine the mixing angles $\alpha_{1,2,3}$. The factor $\mathrm{sign}(R_{23})$ sets the sign of the singlet component in $h_2$, while $R_{13}$ is the singlet admixture of $h_1$. 
		
		\section{Relevant Constraints }
		\label{sec:constraints}
		As with any  BSM extension, the N2HDM parameter space must fulfill extensive theoretical and experimental requirements, which are summarized below. 
		\begin{enumerate}[label=(\roman*)]
			\item Perturbative unitarity is enforced by requiring that all eigenvalues of the tree-level $2\times2$ (pseudo)scalar scattering matrix lie below the perturbative limit $8\pi$ \cite{Muhlleitner:2016mzt}. 
			\item Boundedness From Below (BFB) is ensured by the analytic positivity conditions on the quartic couplings of the N2HDM potential, which guarantee that the scalar potential remains positive along all field directions at large field values \cite{Muhlleitner:2016mzt,Klimenko:1984qx}. The BFB tests are applied by {\tt ScannerS} to every parameter point we consider.	
			\item Vacuum (meta)stability: Vacuum (meta)stability of the EW minimum  is evaluated with the public code {\tt EVADE} \cite{EVADE,Hollik:2018wrr,Ferreira:2019iqb}, which is interfaced to {\tt ScannerS}. If the EW minimum is not the global one at tree level, {\tt EVADE} provides an estimate of the EW vacuum lifetime. Parameter points are retained only if the EW vacuum is global or, if metastable, the estimated lifetime exceeds the age of the Universe. 	
			\item EWPOs: related to the so-called oblique parameters, $S$, $T$ and $U$ \cite{Peskin:1990zt, Peskin:1991sw},  they provide a strong indirect probe of new physics. These parameters, which quantify deviations from the SM in terms of radiative corrections to the $W^\pm$, $Z$ and $\gamma$ self-energies, receive new contributions in the framework of the N2HDM resulting from $h_i$ ($i=1,2,3$), $A$ and $H^\pm$ states. Predictions for $S$, $T$ and $U$ are evaluated at one loop using the implementation in {\tt ScannerS}, for models containing singlet and $SU(2)_L$ doublet scalar sectors \cite{Grimus:2007if, Grimus:2008nb}, and required to be within the $2\sigma$ ellipse of the global fit result \cite{gfitter2018update}.
			\item Flavor-physics observables: these are incorporated using the implementation in {\tt ScannerS}. Since the charged sector of the N2HDM is identical to that of the 2HDM, the corresponding 2HDM bounds can be applied directly to our model for most of the $B$-physics observables. Theoretical predictions for the BRs of $B \to X_s \gamma$, $B_s \to \mu \mu$ and $B_d \to \mu \mu$ are required to be consistent with experimental measurements at $95\%$ CL \cite{Haller:2018nnx}. These constraints impose bounds on $\tan\beta$ and $m_{H^\pm}$ in the N2HDM Type-II and Type-Y.			
			While many additional flavour observables have now been measured with high precision, including semileptonic $b\to s\ell^+\ell^-$ transitions and related angular observables, we do not include them in our fit for two reasons. First, in the CP-conserving N2HDM with a $\mathbb{Z}_2$-symmetric Yukawa sector, tree-level FCNC are absent and the dominant sensitivity to the charged scalar is captured by the channels listed above, which directly constrain $m_{H^\pm}$ and $\tan\beta$. Second, accounting for the current $b\to s\ell^+\ell^-$ anomalies would require additional non-minimal flavour structures that go beyond the purely Higgs-sector interpretation pursued here. We therefore restrict our flavour analysis to the observables that are most constraining for the scenarios under study.
			
			\item Higgs data and direct collider searches: for each viable output of {\tt ScannerS}, we apply collider constraints from LEP, Tevatron  and the LHC using the {\tt HiggsTools} library \cite{Bahl:2022igd}. Direct search limits on additional Higgs bosons are tested with the {\tt HiggsBounds-v.6} subpackage~\cite{Bechtle:2020pkv} at \(\,95\%\) CL, while the 125 GeV signal rate fit is evaluated with the {\tt HiggsSignals-v.3} subpackage~\cite{Bechtle:2020uwn}. We use
			$\Delta\chi^2_{125}=\chi^2_{\rm{N2HDM}} - \chi^2_{\rm{SM}}$ and reject points with $\Delta\chi^2_{125}>6.18$\footnote{The requirement $\Delta\chi^{2}_{125} \le 6.18$ is applied only as a consistency check of the SM-like Higgs sector and is not combined with the 95~GeV and 650~GeV excesses into a global statistical interpretation.}
			, the SM reference being $\chi^2_{\rm{SM}}=152.54$ for 159 degrees of freedom. The required production and decay rates are supplied to {\tt HiggsBounds} and {\tt HiggsSignals} via the {\tt HiggsPredictions} module of {\tt HiggsTools}. In our setup, the cross sections are derived internally via the effective couplings coefficients, while the BRs and total widths are computed with the program {\tt N2HDECAY} \cite{Muhlleitner:2016mzt, Engeln:2018mbg}, interfaced to {\tt ScannerS}, and passed to the analyses through {\tt HiggsPredictions}.  
			
	   \end{enumerate}		
		\section{Explaining the Anomalies}
		\label{sec:anomalies}
		In this section, we briefly summarize the experimental results that can be interpreted as hints for the observed resonances near 95 GeV and 650 GeV being Higgs bosons. The 95 GeV scalar resonance has been observed in multiple channels, starting with a local 2.3$\sigma$ excess reported by LEP in the $e^+e^-\to Z H(\to b\bar{b})$ channel \cite{ALEPH:2006tnd}, consistent with a Higgs boson of mass around 95 GeV and a signal strength of \cite{Cao:2016uwt,Azatov:2012bz}:
		\begin{eqnarray}
			\mu_{b\bar{b}}^{\mathrm{exp}} &=&\frac{\sigma\left( e^+e^- \to Z h_{95} \to Zb\bar{b} \right)} {\sigma^{\mathrm{SM}}\left( e^+e^- \to Z h_{\rm SM}
				\to Z b \bar b \right)}
			\nonumber\\&=& 0.117 \pm 0.057, \label{mubbLEP}
		\end{eqnarray}
		where $h_{95}$ is the lightest CP-even Higgs boson $h_1$ of the N2HDM responsible for the excesses observed around 95 GeV and $h_{\rm SM}$ is interpreted as the SM-like Higgs boson with a mass identical to that of the $h_{95}$.
		The best fits for the LHC di-photon channel observed by CMS and ATLAS for a mass of $\approx 95$ GeV of the Run 2 results of~\cite{CMS:2018cyk,CMS:2024yhz,ATLAS:2024bjr,ATLAS:2023jzc} were combined in~\cite{PhysRevD.109.035005} reaching a local significance of 3.1$\sigma$ with a signal strength given by: 
		\begin{eqnarray}
			\mu_{\gamma\gamma}^{\mathrm{exp}}=\frac{\sigma \left( gg \to h_{95} \to \gamma\gamma \right)}
			{\sigma^{\mathrm{SM}}\left( gg \to h_{\rm SM} \to \gamma\gamma \right)} = 0.24^{+0.09}_{-0.08}. \label{mugagaLHC}
		\end{eqnarray}
		Another anomaly in the low mass region at 95--100 GeV was reported by the CMS collaboration showing a local significance between 2.6$\sigma$ and 3.1$\sigma$ in the di-tau final state. For a mass hypothesis of 95 GeV, the measured signal strength was found to be \cite{CMS:2022goy,Ellwanger:2023zjc}: 
		\begin{eqnarray}
			\mu^{\mathrm{exp}}_{\tau^+\tau^-} = \frac{\sigma (g g \to h_{95} \to \tau^+\tau^-)} {\sigma{^{\mathrm{SM}}}(g g \to h_{\rm SM} \to \tau^+\tau^-)} = 1.38^{+0.69}_{-0.55}. \label{mutautauCMS}
		\end{eqnarray} 
		Additionally, the existence of a Higgs boson at 95 GeV is further supported by a 3.8$\sigma$ local excess (reduced to 2.8$\sigma$ global)  observed in a search by CMS for a 650 GeV resonance decaying into a SM Higgs boson plus a new lighter Higgs boson at 
		90--100 GeV in the $\gamma\gamma b\bar{b}$ final state, using the full Run 2 dataset \cite{CMS:2023boe}. The best fit value for the product of the cross section times the relevant  BRs for this decay channel is given by:
		\begin{eqnarray}
			\sigma^{\rm exp}_{\gamma\gamma b\bar{b}}&=&\sigma(pp \to  X_{650} \to  H_{\rm SM}  Y_{95} \to \gamma\gamma b\bar{b})\nonumber\\ &=& 0.35^{+0.17}_{-0.13}~{\rm fb}.
			\label{sigma650CMS}
		\end{eqnarray}	
		
		As clarified, 
		in our analysis, we focus on the CP-even interpretation of the 650 GeV excess within the N2HDM, taking the resonance to be the heavier CP-even state ($h_3\equiv X_{650}$). In this scenario $h_3$ decays directly into the SM-like Higgs boson ($h_2\equiv H_{\rm SM}$) and the lighter CP-even scalar ($h_1\equiv Y$) near 95 GeV, with $h_1\to b\bar b$. When $h_2 \to \gamma\gamma$, the cascade yields the characteristic $\gamma\gamma b\bar{b}$ final state that we confront with data in the Type-II and Type-Y Yukawa structures. 
		
		\section{Scan and Numerical Results}
		\label{sec:results}
		In what follows, we assess the simultaneous viability of the two hints within the N2HDM, wherein $h_1$ and $h_3$ are identified with the 95 GeV and 650 GeV states, respectively. A comprehensive investigation of this scenario is performed via a random scan over the specified parameter ranges in Tab. \ref{tab:inputs_scan}. We retain only the points that satisfy the theoretical and experimental requirements discussed earlier. The resulting viable set underlies the plots shown below, from which we extract representative BPs with testable predictions for current and upcoming LHC runs.  
		
		\begin{table}[!h]
			\centering
			\renewcommand{\arraystretch}{1.15}
			\setlength{\tabcolsep}{8pt} 
			\begin{adjustbox}{max width=0.5\textwidth}
				\begin{tabular}{cccccc} \Xhline{0.85pt}
					\hline
					\(m_{h_1}\) & \(m_{h_2}\) & \(m_{h_3}\) & \(m_A\) & \(m_{H^\pm}\) & \(\tan\beta\) \\
					\hline
					\([94,\,97]\) & \(125.09\) & \([625,\,675]\) & \([400,\,1000]\) & \([600,\,1000]\) & \([1,\,10]\) \\
					\hline\hline
					\(c^2_{h_2VV}\) & \(c^2_{h_2 t\bar t}\) & \(\mathrm{sign}(R_{23})\) & \(R_{13}\) & \(m_{12}^2\) & \(v_S\) \\
					\hline
					\([0.6,\,1]\) & \([0.6,\,1.2]\) & \(\pm 1\) & \([-1,\,1]\) & \([0,\,5\times 10^5]\) & \([10,\,2000]\) \\ 
					\hline \Xhline{0.85pt}
				\end{tabular}
			\end{adjustbox}
			\caption{Scan ranges for the {\tt ScannerS} input parameters in the N2HDM Type-II and Type-Y. Physical masses and  $v_s$($\mu^2_{12}$) are given in GeV$^{(2)}$.}
			\label{tab:inputs_scan}
		\end{table}
		\begin{figure*}[htpb!]
			\centering
			\includegraphics[width=0.325\textwidth]{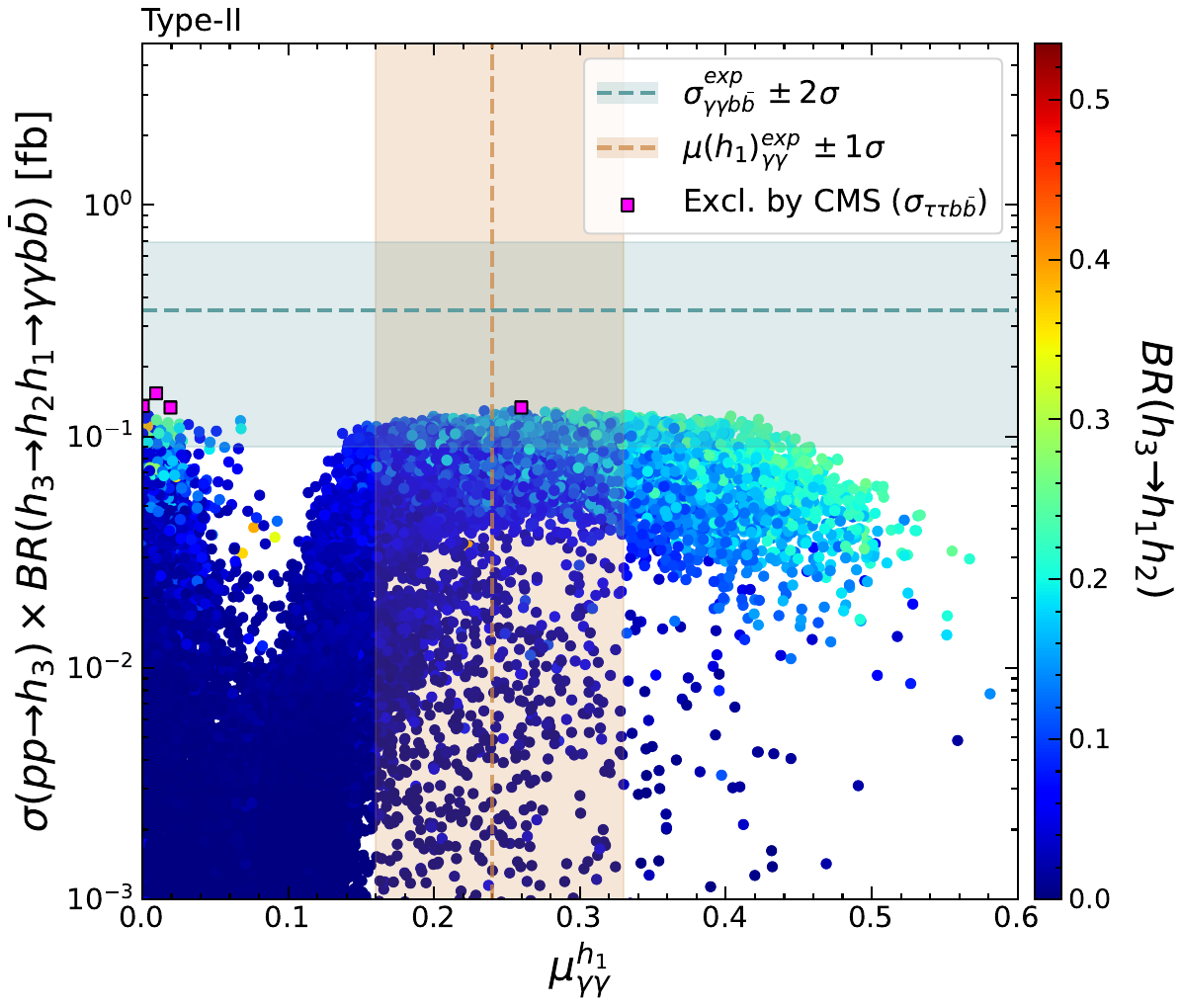}
			\includegraphics[width=0.325\textwidth]{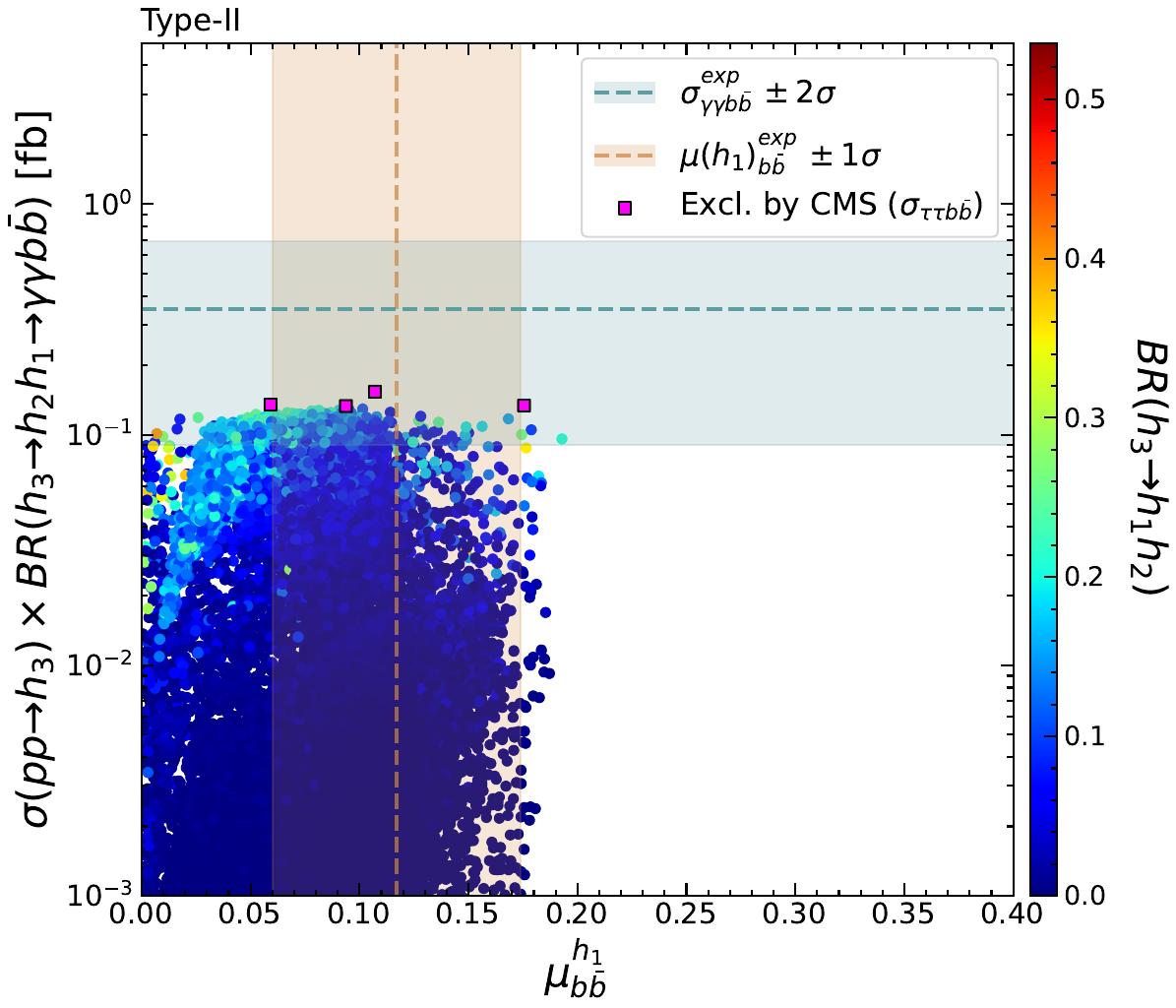}
			\includegraphics[width=0.325\textwidth]{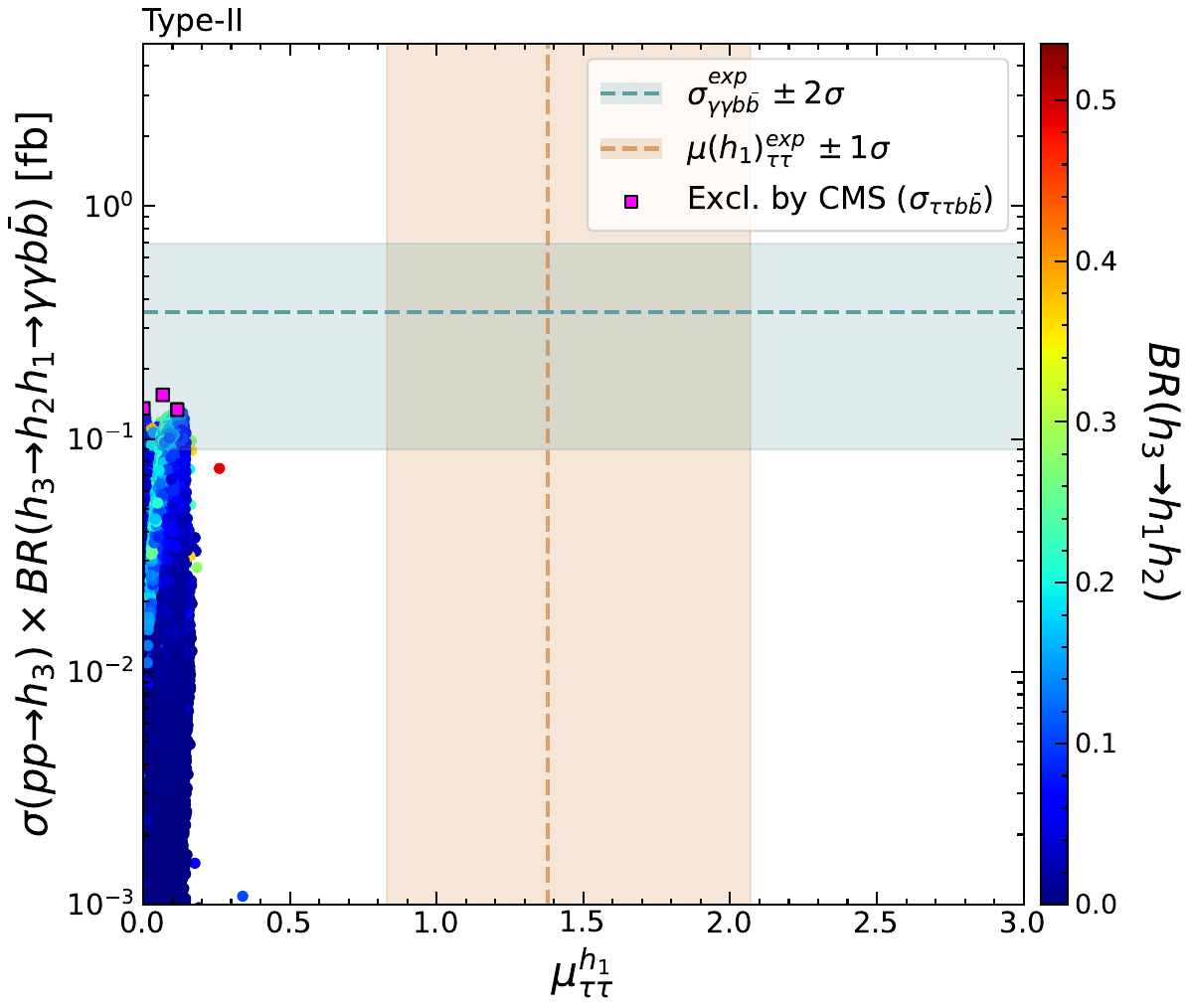}
			\includegraphics[width=0.325\textwidth]{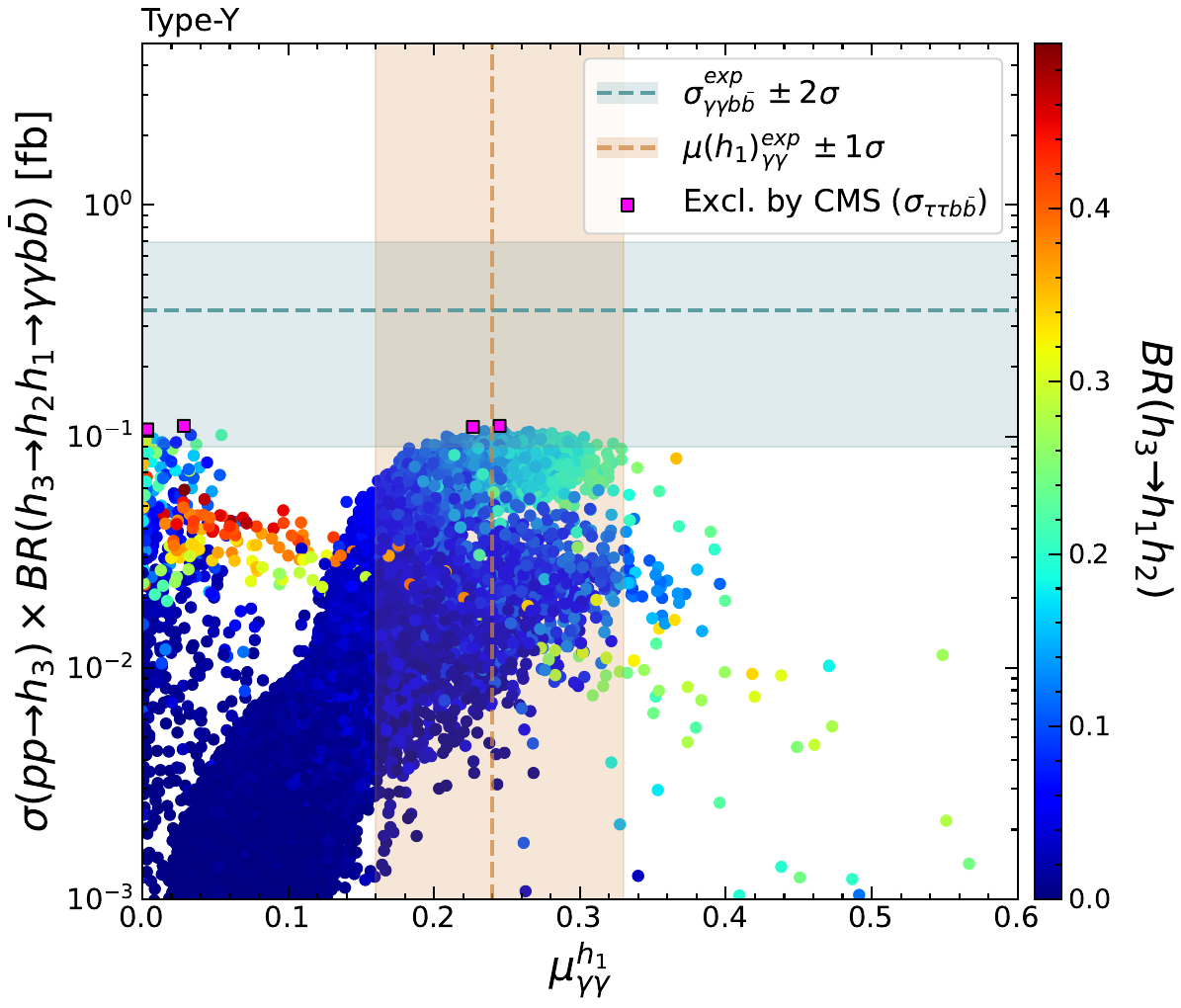}
			\includegraphics[width=0.325\textwidth]{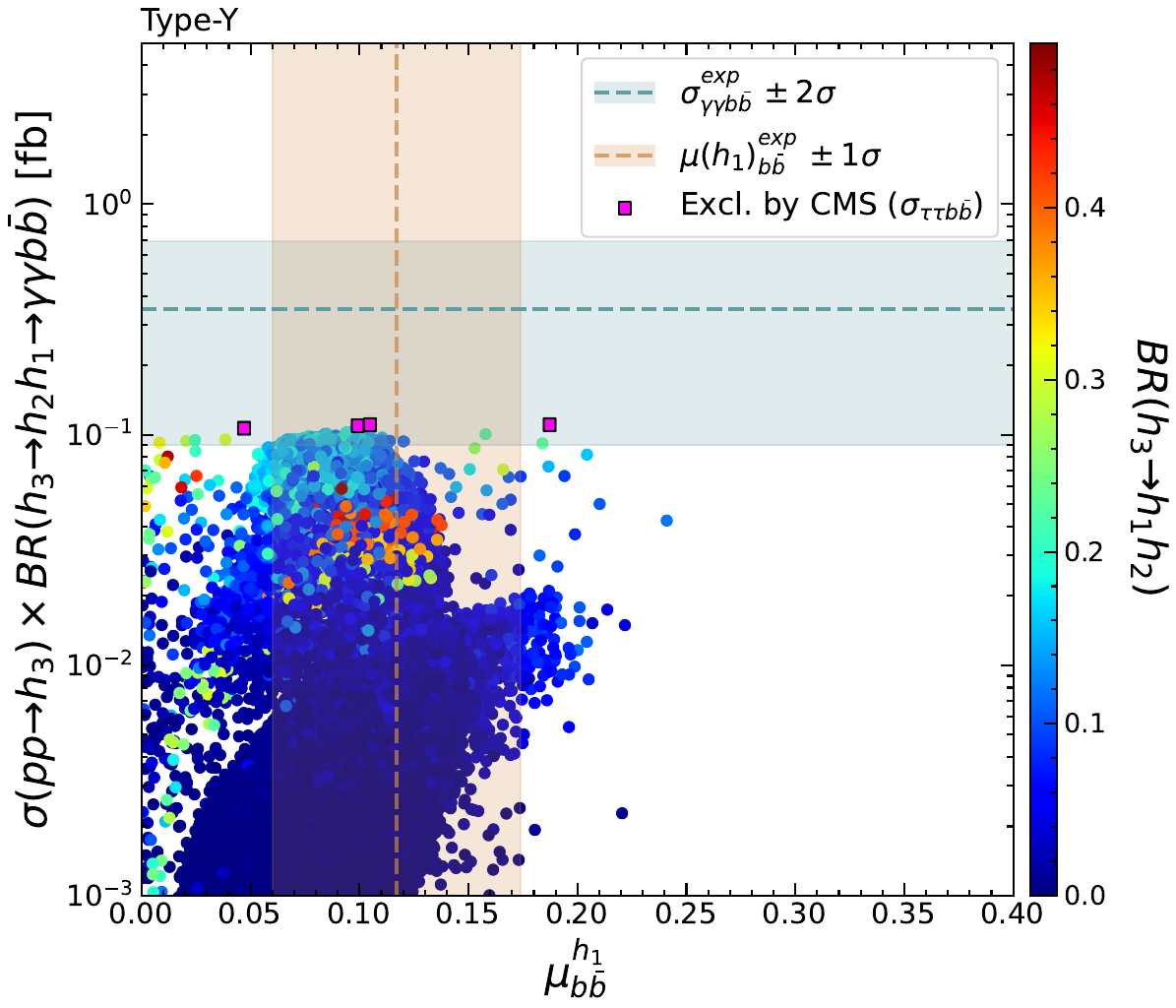}
			\includegraphics[width=0.325\textwidth]{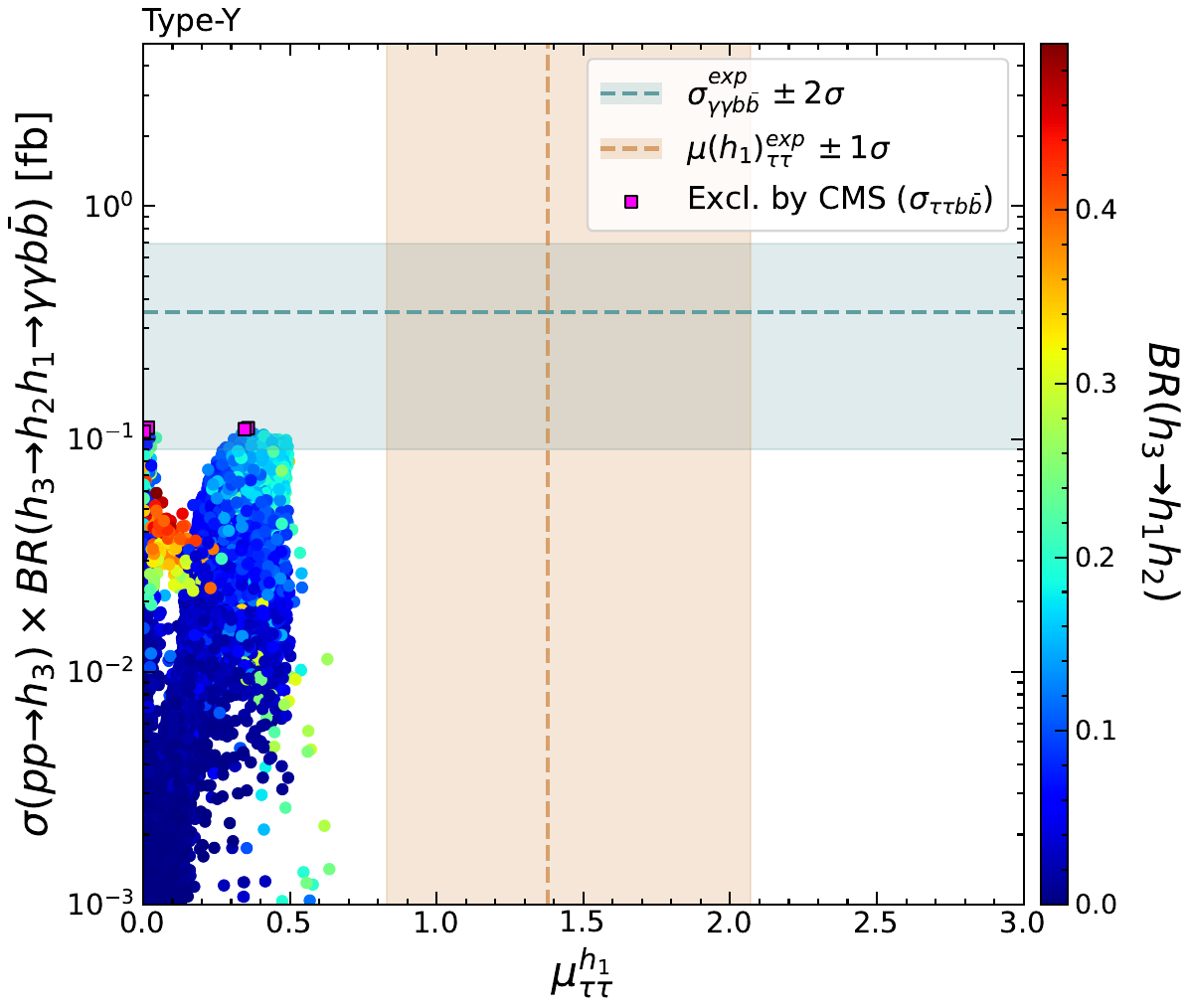}
			\caption{The cross section $\sigma(pp \to h_3)\times {\rm BR}(h_3 \to h_2 h_1 \to \gamma\gamma b\bar{b})$, denoted as $\sigma_{\gamma\gamma b\bar{b}}$, shown as a function of the signal strengths ($\mu^{h_1}_{\gamma\gamma}$, $\mu^{h_1}_{bb}$ and $\mu^{h_1}_{\tau^+\tau^-}$) of the lighter Higgs boson $h_1$ at approximately 95 GeV, within the N2HDM frameworks. The upper(lower) panels correspond to Type-II(Type-Y)  Yukawa interactions. The shaded horizontal blue bands represent the experimentally allowed 2$\sigma$ regions for the cross section of the 650 GeV resonance observed by CMS in the $\gamma\gamma b\bar{b}$ final state. The vertical shaded brown regions represent the experimentally allowed 1$\sigma$ intervals for the signal strengths of the 95 GeV resonance in each decay channel. Colours of the points quantify the ${\rm BR}(h_3 \to h_1 h_2)$. The squares with magenta color indicate points excluded by the CMS Higgs resonance searches in the $\tau^+\tau^- b\bar{b}$ final state~\cite{CMS:2021yci}.}
			\label{fig1}
		\end{figure*}
		To rigorously interpret our predicted cross sections in the $\gamma\gamma b\bar{b}$ channel within the N2HDM, we also rely on the CMS measurements of the closely related $\tau^+\tau^- b\bar{b}$ final state. Specifically, CMS conducted a dedicated analysis at a center-of-mass energy of 13 TeV searching for a heavy Higgs resonance $X_{650}$ decaying into a SM-like Higgs boson $H_{\rm SM}$ and a lighter scalar,  setting, in the absence of any excess, an upper limit of approximately 3 $\rm fb$ at the $95\%$ CL on the production cross section times BRs, $\sigma(gg \to X_{650} \to  H_{\rm SM}  Y_{95} \to \tau^+\tau^- b\bar{b})$, hereafter denoted as $\sigma_{\tau\tau b\bar{b}}$, as clearly outlined in~\cite{CMS:2021yci,Ellwanger:2023zjc}. Given that the rate difference between the $\tau^+\tau^- b\bar{b}$ and $\gamma\gamma b\bar{b}$ final states arises mainly from the differences in the BRs of the SM-like Higgs boson decays to $\tau^+\tau^-$ and $\gamma\gamma$, we adopt the approach described in~\cite{Banik:2023ecr,Ellwanger:2023zjc} by rescaling the experimental limit of 3 $\rm fb$ on the related $\tau^+\tau^- b\bar{b}$ final state. In our N2HDM analysis, we find this ratio of BRs $\left({\rm BR}(H_{\rm SM} \to \tau^+\tau^-)/{\rm BR}(H_{\rm SM} \to \gamma\gamma)\right)$ to be approximately 22.95 for the Type-II framework and about 28.22 for the Type-Y one. Consequently, the experimentally derived limit of 3 $\rm fb$ on $\tau^+\tau^-$ translates into estimated upper limits of about 0.130 $\rm fb$ and 0.106 $\rm fb$ for the $\gamma\gamma b\bar{b}$ final state in Type-II and Type-Y, respectively. These scaled limits provide critical tests for our parameter space exploration, highlighting regions where our model predictions can simultaneously accommodate the observed experimental anomalies around 95 and 650 GeV while remaining consistent with other experimental constraints. 
		Throughout, the signal rates for $pp \to h_3 \to h_2(D) h_1(b\bar{b})$ with $D=\gamma\gamma, \tau^+\tau^-$ at $\sqrt{s}=13$ TeV are evaluated as:
		\begin{equation}
			\sigma(pp \to h_3) \times {\rm BR}(h_3 \to h_2 h_1) \times {\rm BR}(h_2 \to D) \times {\rm BR}(h_1 \to b\bar{b}).	
		\end{equation}
		The production cross section $\sigma(pp \to h_3)$ from gluon fusion (ggF) and $b$-quark fusion (bbF) are computed at Next-to-Next-to-Leading-Order (NNLO) in  QCD using tabulated results from the public program {\tt SusHi} \cite{Harlander:2012pb,Harlander:2016hcx}, while the BRs are obtained with {\tt N2HDECAY}; both codes are interfaced to {\tt ScannerS} for each parameter point.  
		
		%
		\begin{figure*}[htpb!]
			\centering
			\includegraphics[width=0.325\textwidth]{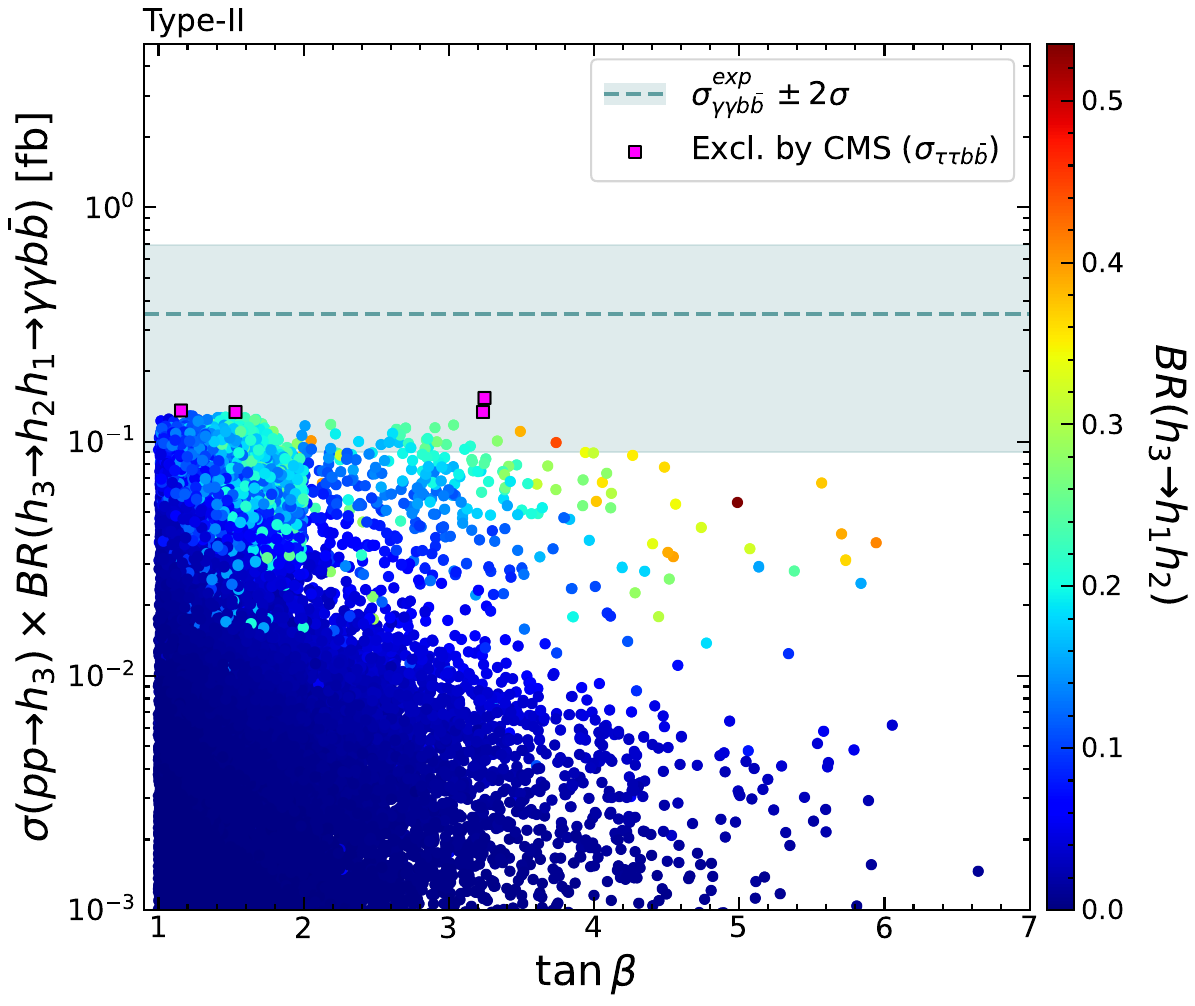}
			\includegraphics[width=0.325\textwidth]{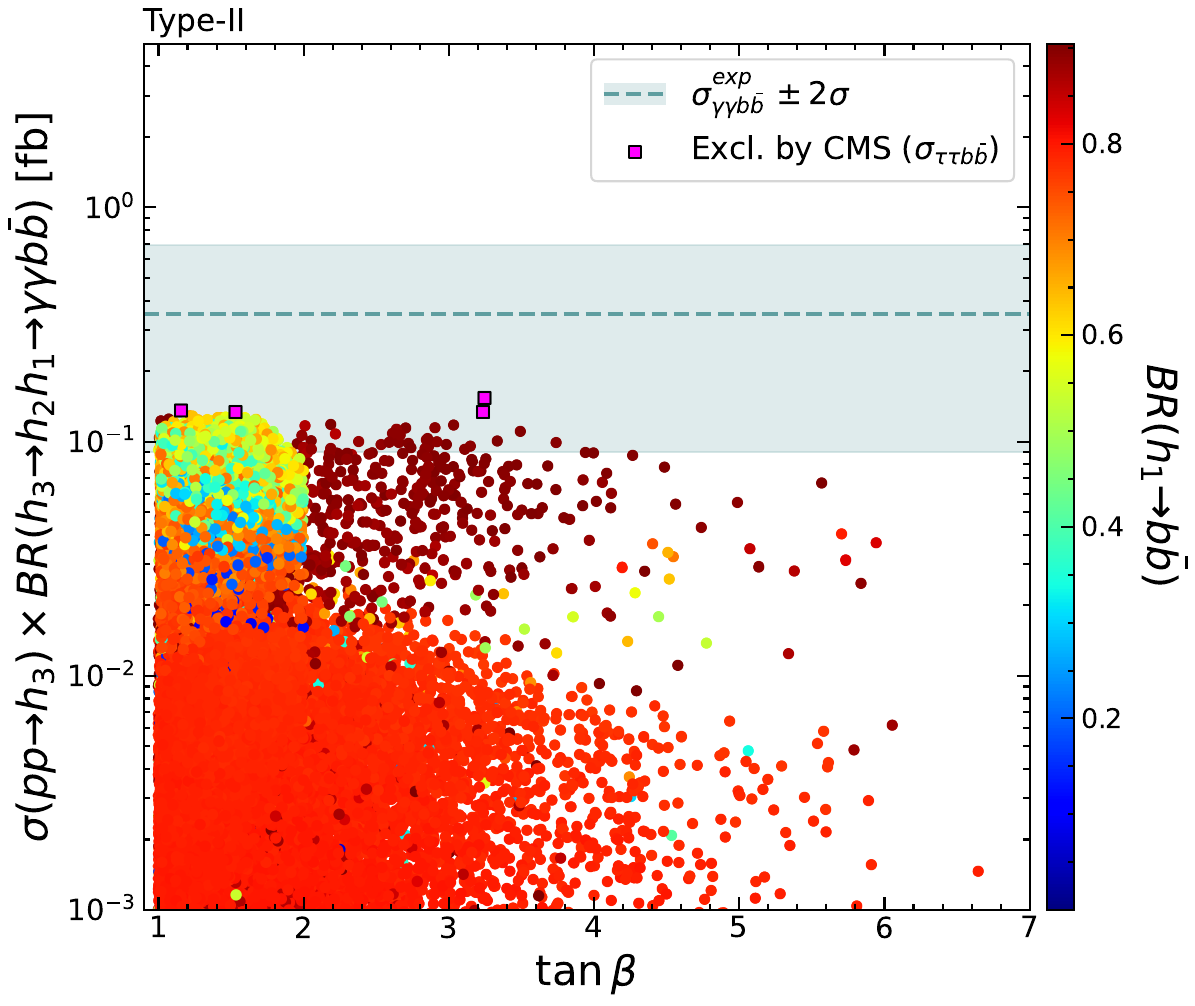}
			\includegraphics[width=0.325\textwidth]{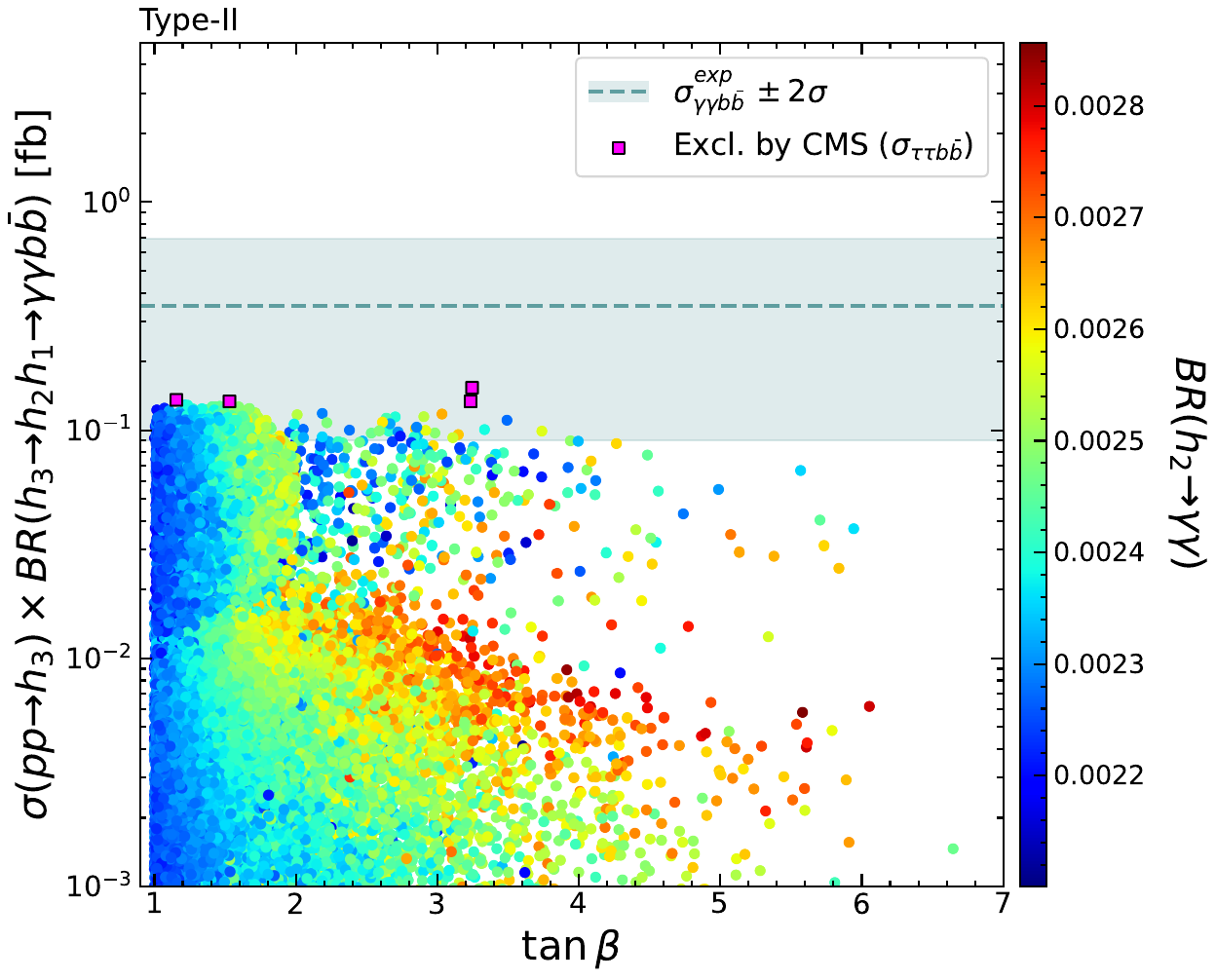}
			\includegraphics[width=0.325\textwidth]{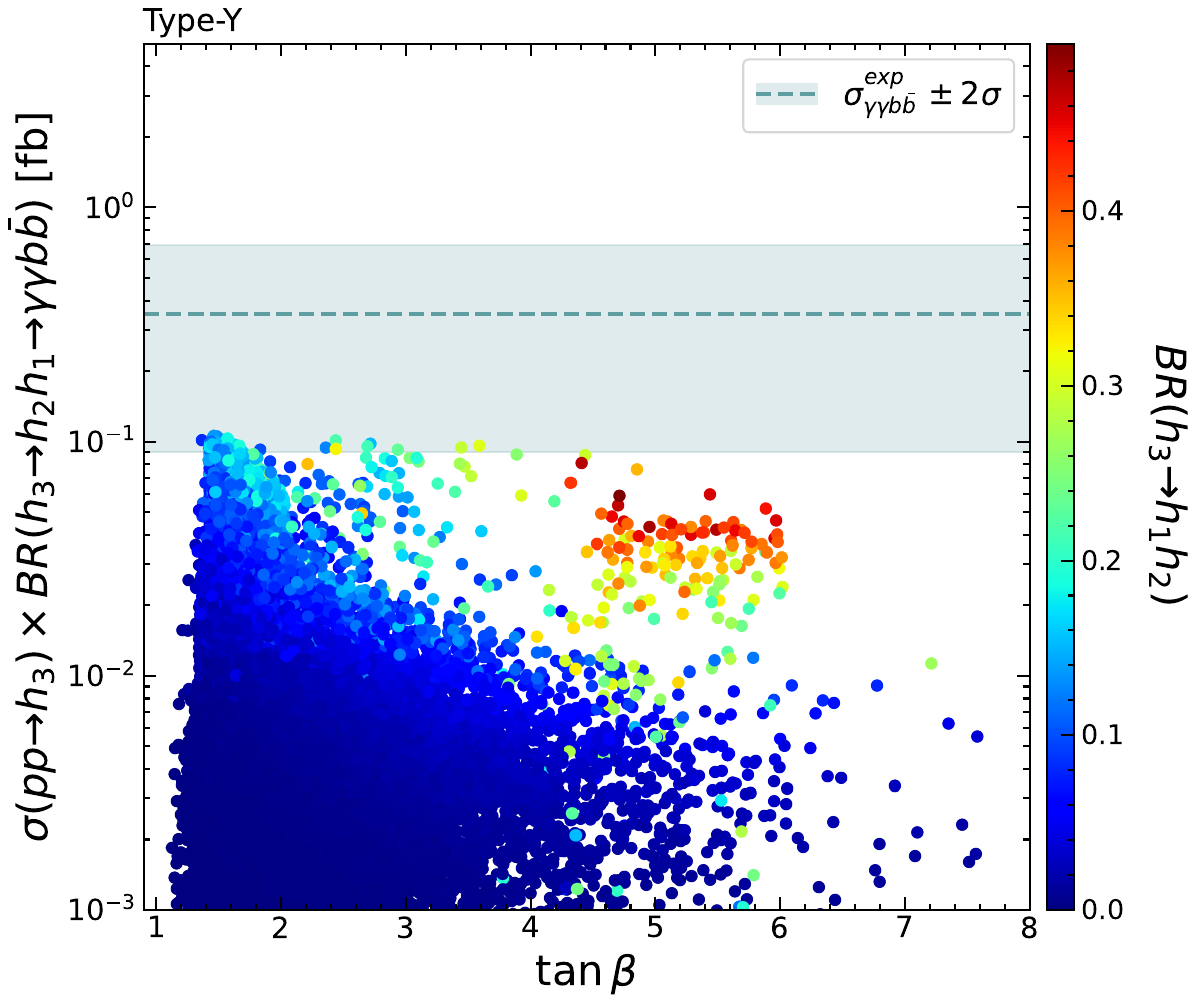}
			\includegraphics[width=0.325\textwidth]{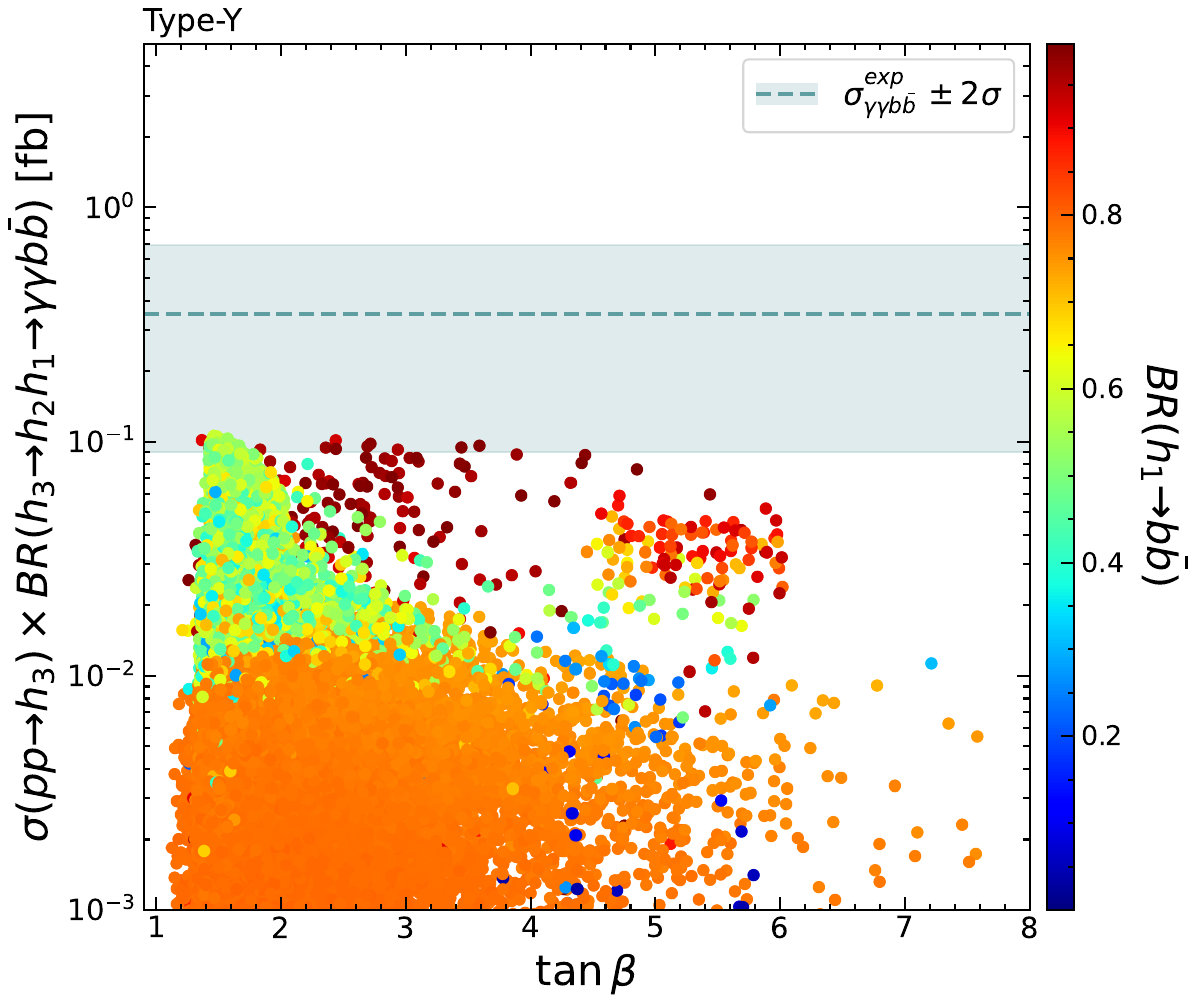}
			\includegraphics[width=0.325\textwidth]{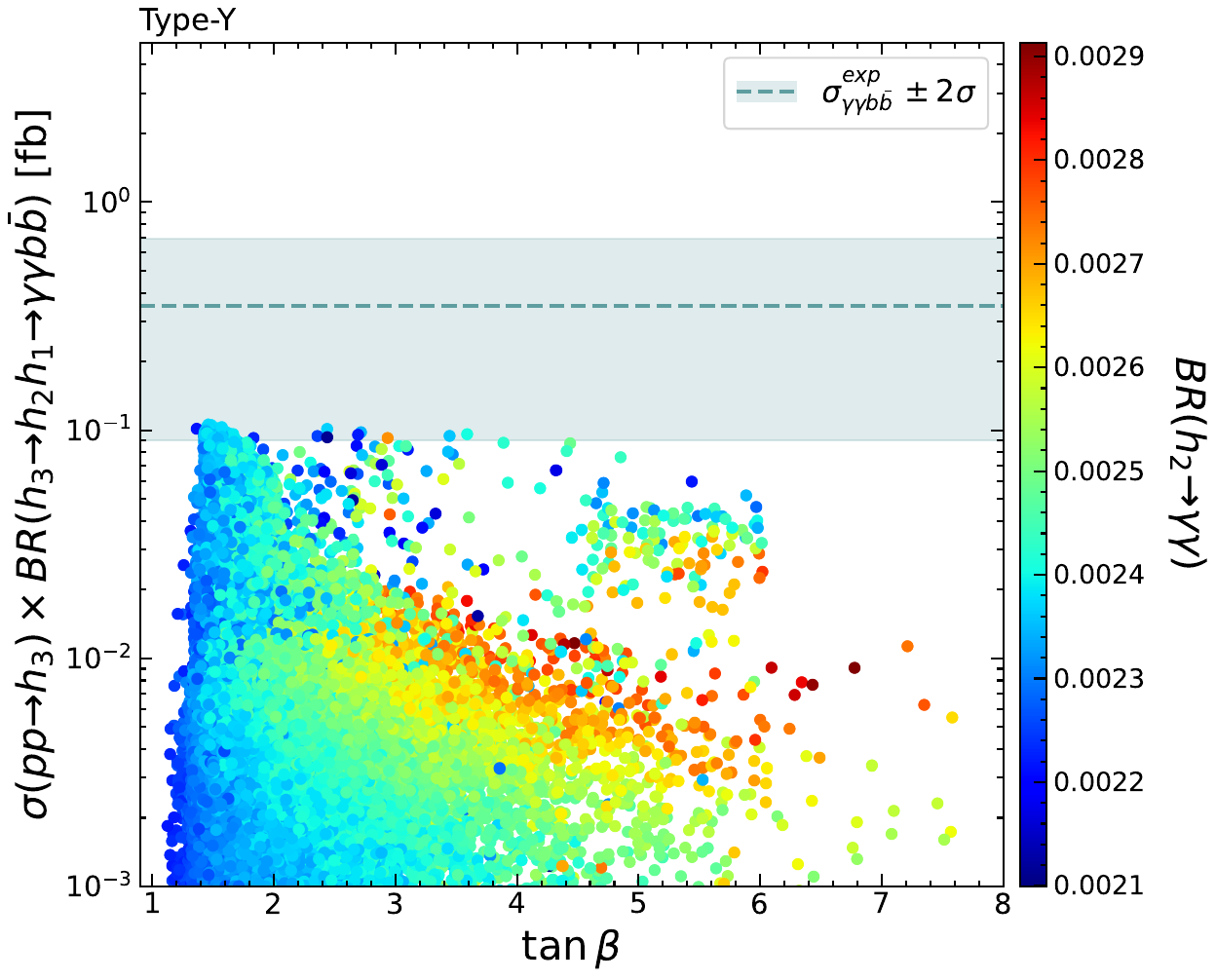}
			\caption{The cross section $\sigma_{\gamma\gamma b\bar{b}}$ as a function of $\tan\beta$ values versus the BRs of $h_3 \to h_1 h_2$ (left panel), $h_1 \to b\bar{b}$ (middle) and $h_2 \to \gamma\gamma$ (right) in the N2HDM Type-II (top) and Type-Y (bottom). Magenta squares are the same as in Fig \ref{fig1}.}
			\label{fig2}
		\end{figure*}
		Fig.~\ref{fig1} illustrates the scatter plots of the predicted $\gamma\gamma b\bar{b}$ rate yield from the CP-even cascade $h_3 \to h_2 h_1$, representing the experimental resonance at 650 GeV, against the signal strengths of the lighter Higgs boson $h_1$ around 95 GeV in the context of the N2HDM Type-II (top panels) and Type-Y (bottom panels). The allowed parameter space corresponds to the intersection of the horizontal shaded bands (blue), indicating the $2\sigma$ cross section range for the 650 GeV resonance, and the vertical shaded bands (brown), representing the $1\sigma$ signal strength intervals for the 95 GeV resonance. Points marked in magenta are excluded by the CMS $\tau^+\tau^- b\bar{b}$ search~\cite{CMS:2021yci}. The color scale encodes ${\rm BR}(h_3 \to h_1 h_2)$. As it can be seen, in the Type-II case (top panels), the scanned points within the overlap form a well-defined region with a clear optimal range within the $1\sigma$ level of $\mu_{\gamma\gamma}^{h_1}$ (left) and $\mu_{b\bar{b}}^{h_1}$ (middle), giving rates up to $\sim 1.2 \times 10^{-1} ~\rm fb$ with ${\rm BR}(h_3 \to h_1 h_2)$ around 20$\%$ for $\mu_{\gamma\gamma}^{h_1} \gtrsim 0.18$ and $\mu_{b\bar{b}}^{h_1} \approx 0.05-0.17$. 
		Similarly, for the Type-Y (bottom panels), the same pattern holds but with a slightly lower reach into the signal region. Thereby, the predicted cross section reaches up to $\sim 1.0 \times 10^{-1} ~\rm fb$ when ${\rm BR}(h_3 \to h_1 h_2) \approx 0.17$, while the most viable points concentrate over the ranges $\mu_{\gamma\gamma}^{h_1} \approx 0.19$--$0.32$ and $\mu_{b\bar{b}}^{h_1} \approx 0.06$--$0.18$. 
		Pushing the $\mu_{\gamma\gamma}^{h_1}$ and $\mu_{b\bar{b}}^{h_1}$ beyond their preferred regions, the cross section decreases significantly, highlighting a limited window of viable parameter space for Type-Y compared with Type-II. Furthermore, turning to leptonic modes, both Yukawa types predict moderate $\mu_{\tau^+\tau^-}$ signal strengths. Points compatible with the CMS band tend to predict $\mu_{\tau^+\tau^-}$ below 0.34(0.63) in Type-II(Type-Y), which is constrained by the exclusion limits based on CMS searches \cite{CMS:2022arx,CMS:2024ulc}, which makes the $\tau^+\tau^-$ signal strength less significant in shaping the simultaneous fit.
		Overall, both Yukawa types can simultaneously reproduce the observed excesses at 95 GeV and 650 GeV within the targeted $2\sigma$ CL, with Type-II reaching slightly larger $\sigma_{\gamma\gamma b\bar{b}}$ than Type-Y. 
		
		Fig.~\ref{fig2} shows the dependence of the cross section $\sigma_{\gamma\gamma b\bar{b}}$ on $\tan\beta$, emphasizing the role of key BRs in both the Type-II (top panels) and Type-Y (bottom panels) realization of the N2HDM. The horizontal shaded band denotes the experimentally allowed $2\sigma$ interval for the predicted rate $\sigma_{\gamma\gamma b\bar{b}}$. Herein, the left panels reveal a significant correlation between $\tan\beta$ and ${\rm BR}(h_3 \to h_1 h_2)$: the allowed points show a clear rise of ${\rm BR}(h_3 \to h_1 h_2)$ with increasing $\tan\beta$ in both Yukawa types. Indeed, for Type-II, the accepted $\tan\beta$ range is from 1.001 to 3.738, with a maximum ${\rm BR}(h_3 \to h_1 h_2)=0.444$ at $\tan\beta=3.738$. For Type-Y, the region is from $\tan\beta \approx 1.364$ to $3.589$,  reaching a maximum ${\rm BR}(h_3 \to h_1 h_2)=0.322$ at $\tan\beta=2.438$.
		This behavior results from the suppression of the dominant decay channel $h_3 \to t\bar{t}$ at higher $\tan\beta$, thereby enhancing $h_3 \to h_1 h_2$. However, at very large $\tan\beta$, the gluon-gluon fusion cross section diminishes due to weakened top-quark loop contributions, limiting the overall enhancement of the signal rate.
		The middle panels show that both Yukawa types predict consistently large values of ${\rm BR}(h_1 \to b\bar{b}) \gtrsim 80\%$, underscoring the dominance of the $b\bar{b}$ final state. At high $\tan\beta$ the production via ggF weakens and bbF cannot compensate, so the rate falls. In Type-II, a few of the highest points are constrained by the CMS $\tau^+\tau^- b\bar{b}$ search, but a  region of viable points remains in the allowed experimental band. 
		The right panels emphasize the role of ${\rm BR}(h_2 \to \gamma\gamma)$, which remains appreciable across the viable parameter space, with slightly larger values in Type-II than in Type-Y, reflecting different sensitivities to loop-induced processes but with modestly altered di-photon rates.
		
		These findings present a coherent picture in
		which a light scalar near 95 GeV and a heavier one 
		around 650 GeV can naturally arise in the N2HDM Type-II and Type-Y
		frameworks and explain well the excesses studied, with the possible exception of the $\tau^+\tau^-$ one for  the lower mass (which is marginal in comparison to $\gamma\gamma$ and $b\bar b$). For such a reason and to enable future tests of the N2HDM, crucially,  in the configuration explaining such anomalies, 
		we provide a selection of four BPs in each Yukawa structure of it, which parameters and corresponding BRs are listed in Tab. \ref{tab:BPs_type2} and Tab. \ref{tab:BRs_type2} for Type-II as well as in Tab. \ref{tab:BPs_type4} and Tab. \ref{tab:BRs_type4} for Type-Y. 
		For all such BPs, the signal strengths for the  $h_1$ state near 95 GeV are fully consistent with experimental LHC and LEP searches at 1$\sigma$, while the predicted 650 GeV $\sigma_{\gamma\gamma b\bar{b}}$ rate falls within the CMS 2$\sigma$ sensitivity. Across these BPs, the heavy CP-even $h_3$ is produced mainly via ggF, couples weakly to $VV$ and decays dominantly to $t\bar{t}$. For moderate values of $\tan\beta$, a sizable $h_3 \to h_2h_1$ decay mode is maintained, which in turn drives the $\sigma_{\gamma\gamma b\bar{b}}$ signal.
		
\begin{figure*}[htpb!]
			\centering
			\includegraphics[width=0.45\textwidth]{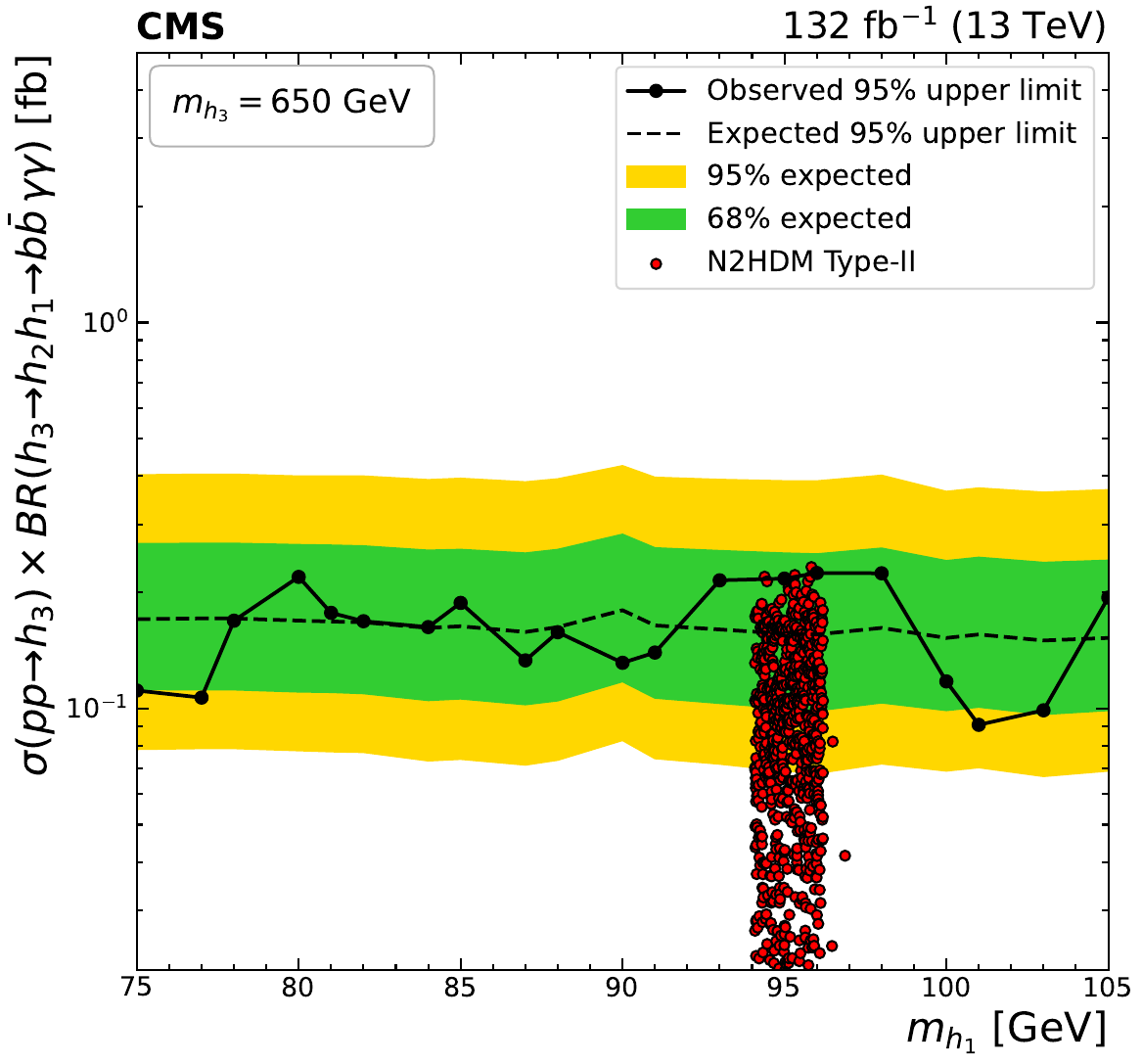}
			\includegraphics[width=0.45\textwidth]{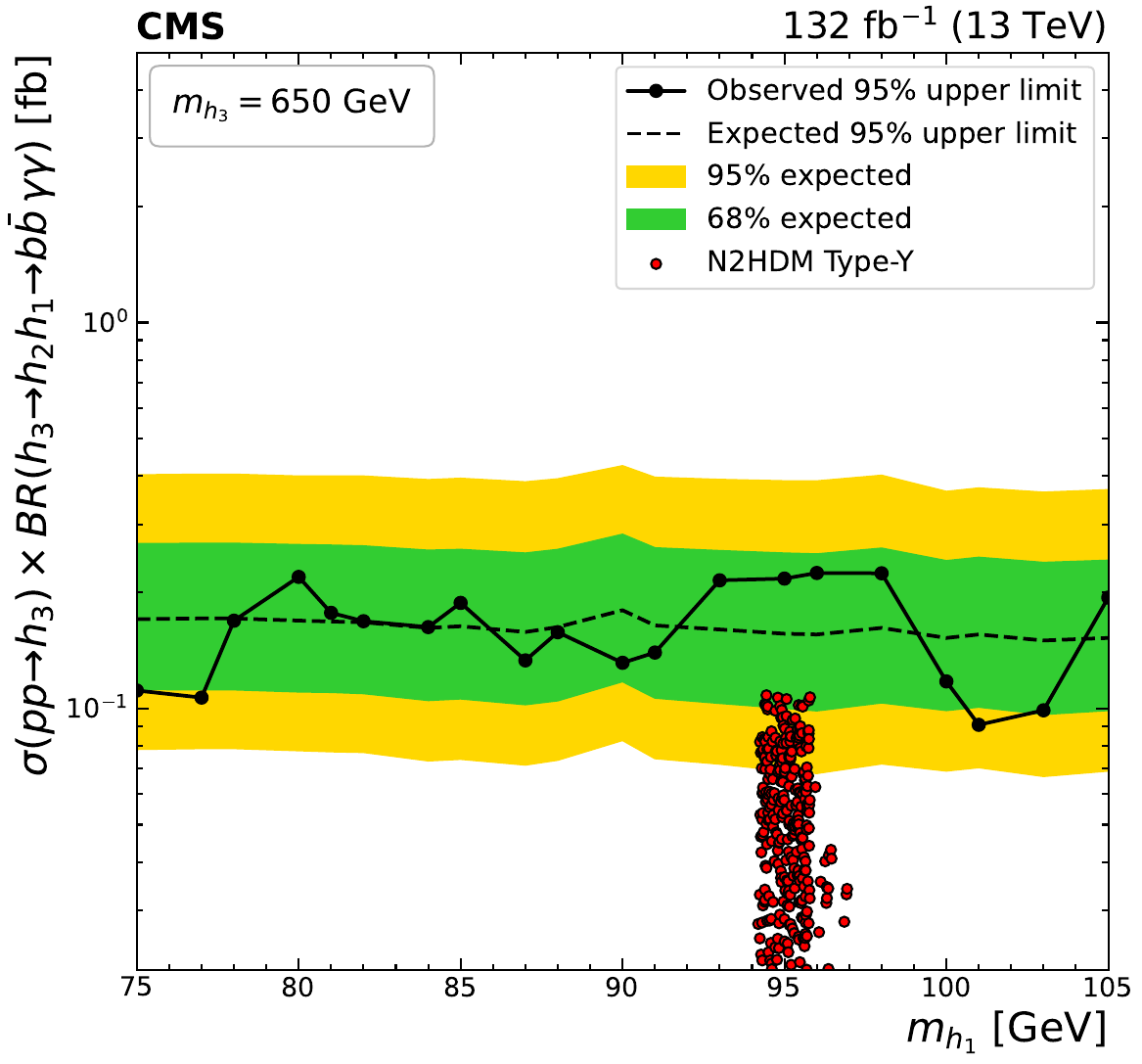}
			\caption{Surviving parameter points for $\sigma(pp \to h_3)\times {\rm BR}(h_3 \to h_2(\to b\bar{b}) h_1(\to \gamma\gamma))$, overlaid onto the observed 95$\%$ CL exclusion limits \cite{CMS:2025qit}. All points satisfy the full set of theoretical and experimental constraints discussed above within the N2HDM Type-II (left) and Type-Y (right).}
			\label{fig3}
		\end{figure*}

\begin{figure*}[htpb!]
			\centering
			\includegraphics[width=0.45\textwidth]{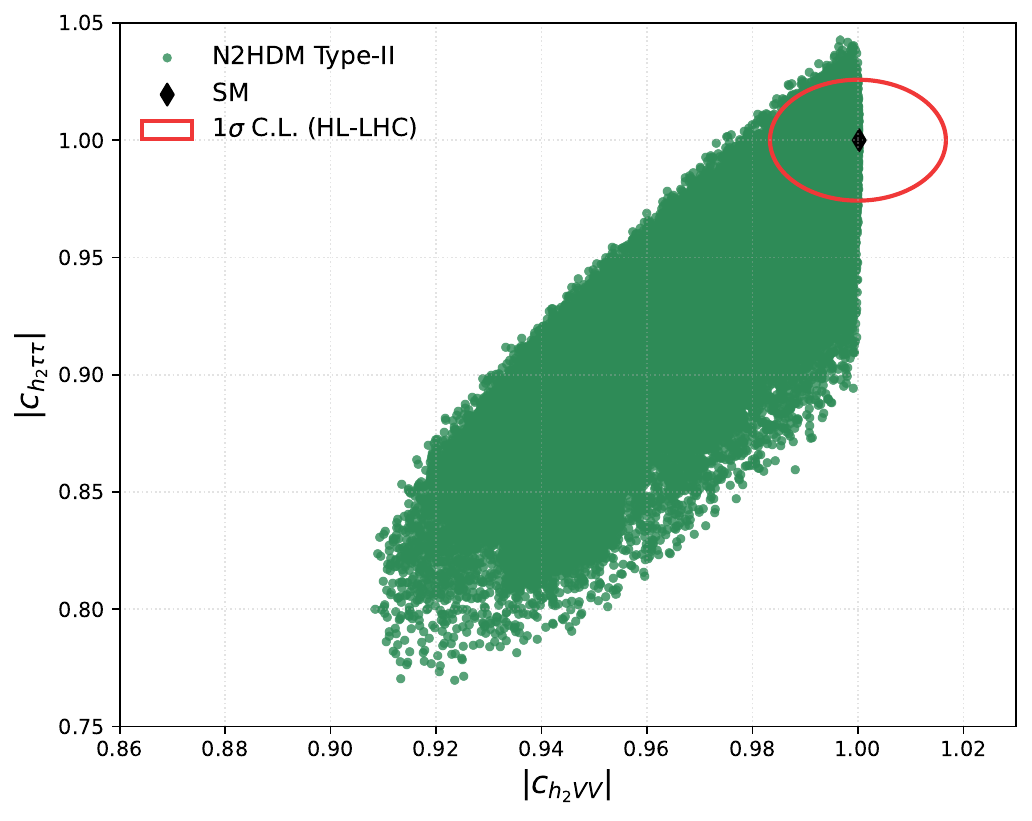}
			\includegraphics[width=0.45\textwidth]{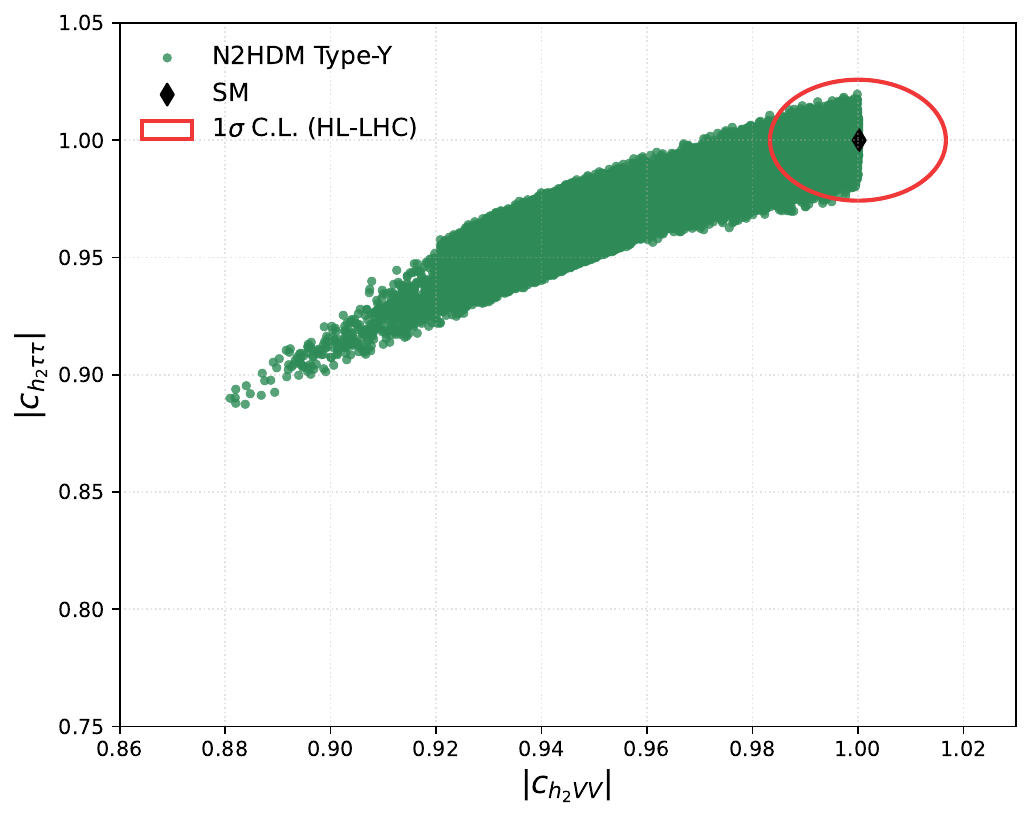} 
			\caption{Projection of our allowed scan points in the ($\lvert c_{h_2VV}\rvert$,$\lvert c_{h_2\tau\tau}\rvert$) plane for the N2HDM Type-II (left), and Type-Y (right). The red ellipse in each panel indicates the projected $1\sigma$ uncertainties at the HL-LHC~\cite{Cepeda:2019klc}.}
			\label{fig4}
		\end{figure*}

		\begin{table}[htpb!]
			\centering
			\renewcommand{\arraystretch}{1.15}  
			\setlength{\tabcolsep}{5pt} 
			\begin{adjustbox}{max width=0.5\textwidth}
				\begin{tabular}{l|cccc} 	\Xhline{0.85pt}
					& \multicolumn{4}{c}{\textbf{Type-II}}  \\
					\Xhline{0.85pt}
					\textbf{Parameters} & BP1 & BP2 & BP3 & BP4   \\
					\Xhline{0.85pt}
					$m_{h_1}$  & 95.28      & 95.81   & 95.03   & 95.40       \\
					$m_{h_2}$  & 125.09    & 125.09    & 125.09    & 125.09     \\
					$m_{h_3}$  & 651.05    & 650.93    & 650.89    & 650.69     \\
					$m_A$      & 717.89    & 717.92     & 709.59    & 653.39     \\
					$m_{H^\pm}$ & 633.30    & 700.23   & 687.13  & 697.28       \\
					$\tan\beta$   & 1.470    & 1.478     & 1.42    & 1.45    \\
					$\alpha_{1}$   & 1.243    & 1.272   & 1.164    & 1.248       \\
					$\alpha_{2}$   & -1.231    & -1.252   & -1.206  & -1.251     \\
					$\alpha_{3}$   & -1.359    & -1.331    & -1.438  & -1.345     \\
					$m_{12}^2$   & 139691.27    & 150158.52 & 124611.04 & 134830.86    \\
					$v_s$       & 40.03    & 41.49     & 40.95  & 48.03     \\
					$\mu^{h_1}_{b\bar{b}}$  & 0.078    & 0.063   & 0.104    & 0.068   \\
					$\mu^{h_1}_{\gamma\gamma}$     & 0.248    & 0.241   & 0.221    & 0.223    \\
					$\sigma_{\gamma\gamma b\bar{b}}~[{\rm fb}] $    & $12.17\times 10^{-2}$    & $11.32\times 10^{-2}$  & $10.01\times 10^{-2}$    & $9.22\times 10^{-2}$    \\
					$\sigma_{\tau\tau b\bar{b}}~[{\rm fb}] $   & 3.035    & 2.860   & 2.437   & 2.354   \\
					\Xhline{0.85pt}
				\end{tabular}
			\end{adjustbox}
			\caption{BPs for the Type-II Yukawa framework.}
			\label{tab:BPs_type2}
		\end{table}
		\begin{table}[htpb!]
			\centering
			\renewcommand{\arraystretch}{1.15}  
			\setlength{\tabcolsep}{5pt} 
			\begin{adjustbox}{max width=0.5\textwidth}
				\begin{tabular}{l|cccc} 	\Xhline{0.85pt}
					& \multicolumn{4}{c}{\textbf{Type-Y}}  \\
					\Xhline{0.85pt}
					\textbf{Parameters} & BP1 & BP2 & BP3 & BP4   \\
					\Xhline{0.85pt}
					$m_{h_1}$  & 95.48      & 95.06   & 95.75   & 95.43       \\
					$m_{h_2}$  & 125.09    & 125.09    & 125.09    & 125.09       \\
					$m_{h_3}$  & 650.49    & 649.41    & 649.89    & 650.42     \\
					$m_A$     & 661.74    & 657.25     & 664.44    & 656.09   \\
					$m_{H^\pm}$  & 681.22    & 686.44   & 710.88  & 670.71       \\
					$\tan\beta$  & 1.50   & 1.49     & 1.50    & 1.57      \\
					$\alpha_{1}$   & 1.225    & 1.227   & 1.238    & 1.233       \\
					$\alpha_{2}$   & -1.176    & -1.206   & -1.194    & -1.184  \\
					$\alpha_{3}$   & -1.385    & -1.336   & -1.377   & -1.393   \\
					$\mu_{12}$   & 169235.35   & 169329.11  & 167695.59  & 166901.55    \\
					$v_s$    & 106.92    & 104.47     & 104.19    & 103.56     \\
					$\mu^{h_1}_{b\bar{b}}$    & 0.096    & 0.070   & 0.084    & 0.094   \\
					$\mu^{h_1}_{\gamma\gamma}$  & 0.268    & 0.255   & 0.253  & 0.256    \\
					$\sigma_{\gamma\gamma b\bar{b}}~[{\rm fb}] $    & $9.94\times 10^{-2}$    & $9.38\times 10^{-2}$    & $9.23\times 10^{-2}$    & $9.01\times 10^{-2}$    \\
					$\sigma_{\tau\tau b\bar{b}}~[{\rm fb}] $   & 2.991    & 2.808   & 2.814    & 2.673   \\
					\Xhline{0.85pt}
				\end{tabular}
			\end{adjustbox}
			\caption{BPs for the Type-Y Yukawa framework.}
			\label{tab:BPs_type4}
		\end{table}
		\begin{table}[htpb!]
			\centering
			\begin{adjustbox}{max width=0.5\textwidth}
				\begin{tabular}{@{}lcccccc@{}}
					\Xhline{0.85pt}
					& \multicolumn{6}{c}{\textbf{BRs in Type-II}} \\
					\Xhline{0.85pt}
					$h_1$ & $b\bar{b}$ & $c\bar{c}$ & $\tau^+\tau^-$ & $WW$ & $ZZ$ & $\gamma\gamma$ \\
					\Xhline{0.85pt}
					BP1 & $0.611$ & $0.118$ & $0.061$ & $1.0\times 10^{-2}$ & $1.40\times 10^{-3}$ & $2.57\times 10^{-3}$ \\
					BP2 & $0.570$ & $0.133$ & $0.057$ & $1.19\times 10^{-2}$ & $1.61\times 10^{-3}$ & $2.78\times 10^{-3}$ \\
					BP3 & $0.688$ & $0.089$ & $0.069$ & $0.77\times 10^{-2}$ & $1.10\times 10^{-3}$ & $2.08\times 10^{-3}$ \\
					BP4 & $0.599$ & $0.123$ & $0.060$ & $1.04\times 10^{-2}$ & $1.45\times 10^{-3}$ & $2.57\times 10^{-3}$ \\
					\hline
					$h_2$ & $b\bar{b}$ & $c\bar{c}$ & $\tau^+\tau^-$ & $WW$ & $ZZ$ & $\gamma\gamma$ \\
					\hline
					BP1 & $0.557$ & $0.033$ & $0.059$ & $0.226$ & $0.028$ & $2.40\times 10^{-3}$ \\
					BP2 & $0.558$ & $0.033$ & $0.059$ & $0.226$ & $0.028$ & $2.37\times 10^{-3}$ \\
					BP3 & $0.544$ & $0.035$ & $0.058$ & $0.233$ & $0.029$ & $2.40\times 10^{-3}$ \\
					BP4 & $0.559$ & $0.033$ & $0.060$ & $0.225$ & $0.028$ & $2.35\times 10^{-3}$ \\
					\hline
					$h_3$ & $t\bar{t}$ & $h_1 h_1$ & $h_1 h_2$ & $h_2 h_2$ & $WW$ & $ZZ$ \\
					\hline
					BP1 & $0.460$ & $0.329$ & $0.171$ & $4.49\times 10^{-3}$ & $0.021$ & $0.010$ \\
					BP2 & $0.443$ &  $0.344$ & $0.173$ & $7.02\times 10^{-3}$ & $0.019$ & $9.47\times 10^{-3}$ \\ 
					BP3 & $0.631$ & $0.17$ & $0.121$ & $9.42\times 10^{-3}$ & $0.038$ & $0.018$ \\
					BP4 & $0.553$ &  $0.275$ & $0.129$ & $6.41\times 10^{-3}$ & $0.022$ & $0.010$ \\
					\hline
					$A$ & $t\bar{t}$ & $b\bar{b}$ & $gg$ & $Zh_1$ & $Zh_2$ & $Zh_3$ \\
					\hline
					BP1 & $0.927$ & $1.30\times 10^{-3}$ & $2.34\times 10^{-3}$ & $0.051$ & $0.012$ & $1.10\times 10^{-4}$ \\
					BP2 & $0.929$ & $1.33\times 10^{-3}$ & $2.34\times 10^{-3}$ & $0.055$ & $0.011$ & $1.11\times 10^{-4}$ \\
					BP3 & $0.943$ & $1.17\times 10^{-3}$ & $2.40\times 10^{-3}$ & $0.032$ & $0.021$ & $5.05\times 10^{-5}$ \\
					BP4 & $0.947$ & $1.28\times 10^{-3}$ & $2.58\times 10^{-3}$ & $0.039$ & $0.008$ & $1.00\times 10^{-11}$ \\
					\hline
					$H^\pm$ & $tb$ & $ts$ & $\tau\nu$ & $Wh_1$ & $Wh_2$ & $Wh_3$ \\
					\hline
					BP1 & $0.946$ & $1.54\times 10^{-3}$ & $2.05\times 10^{-4}$ & $0.042$ & $0.009$ & $0$ \\
					BP2 & $0.931$ & $1.52\times 10^{-3}$ & $2.05\times 10^{-4}$ & $0.055$ & $0.011$ & $2.85\times 10^{-5}$ \\
					BP3 & $0.945$ & $1.54\times 10^{-3}$ & $1.80\times 10^{-4}$ & $0.031$ & $0.020$ & $5.33\times 10^{-6}$ \\
					BP4 & $0.939$ & $1.53\times 10^{-3}$ & $1.93\times 10^{-4}$ & $0.048$ & $0.010$ & $2.04\times 10^{-5}$ \\
					\Xhline{0.85pt}
				\end{tabular}
			\end{adjustbox}
			\caption{BRs for the Type-II BPs.}
			\label{tab:BRs_type2}
		\end{table}
		\begin{table}[htpb!]
			\centering
			\begin{adjustbox}{max width=0.5\textwidth}
				\begin{tabular}{@{}lcccccc@{}}
					\Xhline{0.85pt}
					& \multicolumn{6}{c}{\textbf{BRs in Type-Y}} \\
					\Xhline{0.85pt}
					$h_1$ & $b\bar{b}$ & $c\bar{c}$ & $\tau^+\tau^-$ & $WW$ & $ZZ$ & $\gamma\gamma$ \\
					\hline
					BP1 & $0.553$ & $0.092$ & $0.191$ & $8.38\times 10^{-3}$ & $1.16\times 10^{-3}$ & $2.14\times 10^{-3}$ \\
					BP2 & $0.479$ & $0.107$ & $0.223$ & $8.60\times 10^{-3}$ & $1.23\times 10^{-3}$ & $2.29\times 10^{-3}$ \\
					BP3 & $0.535$ & $0.095$ & $0.199$ & $0.01\times 10^{-3}$ & $1.22\times 10^{-3}$ & $2.19\times 10^{-3}$ \\
					BP4 & $0.565$ & $0.089$ & $0.186$ & $8.34\times 10^{-3}$ & $1.15\times 10^{-3}$ & $2.18\times 10^{-3}$ \\
					\hline
					$h_2$ & $b\bar{b}$ & $c\bar{c}$ & $\tau^+\tau^-$ & $WW$ & $ZZ$ & $\gamma\gamma$ \\
					\hline
					BP1 & $0.553$ & $0.032$ & $0.071$ & $0.222$ & $0.027$ & $2.37\times 10^{-3}$ \\
					BP2 & $0.557$ & $0.032$ & $0.070$ & $0.221$ & $0.027$ & $2.36\times 10^{-3}$ \\
					BP3 & $0.549$ & $0.032$ & $0.072$ & $0.224$ & $0.028$ & $2.38\times 10^{-3}$ \\
					BP4 & $0.556$ & $0.032$ & $0.070$ & $0.221$ & $0.027$ & $2.37\times 10^{-3}$ \\
					\hline
					$h_3$ & $t\bar{t}$ & $h_1 h_1$ & $h_1 h_2$ & $h_2 h_2$ & $WW$ & $ZZ$ \\
					\hline
					BP1 & $0.716$ & $0.037$ & $0.163$ & $0.026$ & $0.035$ & $0.017$ \\
					BP2 & $0.686$ &  $0.065$ & $0.175$ & $0.022$ & $0.031$ & $0.015$ \\
					BP3 & $0.705$ & $0.044$ & $0.159$ & $0.031$ & $0.037$ & $0.018$ \\
					BP4 & $0.727$ &  $0.041$ & $0.157$ & $0.023$ & $0.031$ & $0.015$ \\
					\hline
					$A$ & $t\bar{t}$ & $b\bar{b}$ & $gg$ & $Zh_1$ & $Zh_2$ & $Zh_3$ \\
					\hline
					BP1 & $0.938$ & $1.46\times 10^{-3}$ & $2.54\times 10^{-3}$ & $0.048$ & $8.13\times 10^{-3}$ & $1.30\times 10^{-8}$ \\
					BP2 & $0.931$ & $1.43\times 10^{-3}$ & $2.53\times 10^{-3}$ & $0.057$ & $6.62\times 10^{-3}$ & $2.13\times 10^{-9}$ \\
					BP3 & $0.936$ & $1.48\times 10^{-3}$ & $2.52\times 10^{-3}$ & $0.049$ & $9.97\times 10^{-3}$ & $4.71\times 10^{-8}$ \\
					BP4 & $0.943$ & $1.78\times 10^{-3}$ & $2.57\times 10^{-3}$ & $0.045$ & $7.04\times 10^{-3}$ & $4.75\times 10^{-10}$ \\
					\hline
					$H^\pm$ & $tb$ & $ts$ & $\tau\nu$ & $Wh_1$ & $Wh_2$ & $Wh_3$ \\
					\hline
					BP1 & $0.934$ & $1.52\times 10^{-3}$ & $4.32\times 10^{-5}$ & $0.054$ & $0.009$ & $2.52\times 10^{-6}$ \\
					BP2 & $0.924$ & $1.50\times 10^{-3}$ & $4.27\times 10^{-5}$ & $0.066$ & $0.007$ & $6.44\times 10^{-6}$ \\
					BP3 & $0.926$ & $1.51\times 10^{-3}$ & $4.28\times 10^{-5}$ & $0.059$ & $0.012$ & $9.91\times 10^{-5}$ \\
					BP4 & $0.940$ & $1.53\times 10^{-3}$ & $4.35\times 10^{-5}$ & $0.050$ & $0.007$ & $3.47\times 10^{-7}$ \\
					\Xhline{0.85pt}
				\end{tabular}
			\end{adjustbox}
			\caption{BRs for the Type-Y BPs.}
			\label{tab:BRs_type4}
		\end{table}
		Complementing the foregoing analysis, the CMS collaboration has recently released the results of an analysis for a possible heavy spin-0 resonance in the inverted final state $h_3 \to h_2(\to b\bar{b}) h_1(\to \gamma\gamma)$ \cite{CMS:2025qit}. We present in Fig.~\ref{fig3} our N2HDM predictions Type-II (left panel) and Type-Y (right panel) superimposed onto the observed 95$\%$ CL exclusion limits from this CMS $b\bar{b} \gamma\gamma$ search \cite{CMS:2025qit}. All surviving points lie below the exclusion contours for both types, thereby confirming consistency with the CMS sensitivity band. Additionally, CMS has also tested the possibility of an
		excess from the process $pp \to X_{650} \to Y_{95}(\to \gamma\gamma) H_{\rm SM}(\to \tau\tau)$  setting an observed 95$\%$ CL upper limit of 1.6208 $\rm fb$ \cite{CMS:2025tqi}. Our N2HDM predictions are much smaller: $\sigma(pp \to h_3 \to h_1(\to \gamma\gamma)h_2(\to \tau\tau)) \approx 2.66 \times 10^{-2}$ $\rm fb$ (Type-II) and around $1.75 \times 10^{-2}$ $\rm fb$ (Type-Y). Thus, this channel lies significantly below the current sensitivity by factors of $\sim60$ to $92$, showing no anomaly and leaving our viable parameter space unconstrained. Overall, the recent $b\bar{b} \gamma\gamma$ and $\gamma\gamma \tau\tau$ bounds are fully compatible with our interpretation, and provide promising additional  probes of the N2HDM Type-II and Type-Y in, possibly,  Run-3 and, more probably, at the HL-LHC~\cite{Cepeda:2019klc}. 
		Taken together, the existing $\gamma\gamma b\bar b$, $\tau^+\tau^- b\bar b$, $b\bar b\,\gamma\gamma$ and $\gamma\gamma\tau^+\tau^-$ searches already probe the correlated cascade structure in several complementary ways, and the same channels at Run~3 and especially at the HL-LHC will be able either to confirm the present picture or to close the remaining window of N2HDM parameter space identified in our scan. As depicted in Fig.~\ref{fig4}, the HL-LHC projection is displayed in the plane of the Higgs normalized couplings ($\lvert c_{h_2VV}\rvert$,$\lvert c_{h_2\tau\tau}\rvert$), defined relative to their SM values. 
		Our surviving scan points compatible with the observed excesses are superimposed. A subset of these points lies within the projected HL-LHC $1\sigma$ sensitivity contour for these reduced couplings, indicating that part of the excess-motivated parameter space could remain viable with future precision Higgs measurements.
		
		\section{Conclusions}
		\label{sec:conlusion}
		In this paper, we have shown how the N2HDM can successfully be invoked to explain a series of anomalies in LEP (from ADLO)  and LHC (from  ATLAS and CMS) data, which can be attributed to an extended Higgs sector. In fact, excesses have been found at around 95 GeV at both colliders, in the former case for  the $b\bar b$ channel while in the latter case for the $\gamma\gamma$ and $\tau^+\tau^-$ 
		channels. Furthermore, the CMS collaboration has also published results on anomalous events in the $\gamma\gamma b\bar b$ channels clustering around 650 GeV. The N2HDM provides three neutral CP-even Higgs states ($h_{1,2,3}$) and one neutral CP-odd Higgs state ($A$) (alongside two charged Higgses ($H^\pm$)). These can be arranged in mass and coupling spectra of the N2HDM in such a way that $h_2$ is the SM-like Higgs state with mass 125 GeV (the one discovered at CERN in 2012), $h_1$  has a mass of 95 GeV and either the $h_3$ or the $A$ has a mass of 650 GeV. In such circumstances, the LEP and LHC anomalies at  95 GeV can be addressed through $e^+e^-\to Z^*\to Z h_1(\to b\bar b)$ and $gg\to h_1(\to \gamma\gamma$, $b\bar b)$ production and decay, respectively, while the LHC one at  650 GeV can be explained through either $gg\to h_3\to h_2(\to\gamma\gamma)h_1(\to b\bar b)$ or $gg\to A\to h_2(\to\gamma\gamma) Z(\to b\bar b)$ (recall that the mass difference $m_{h_1}-m_Z$ is smaller than the resolution of the $b\bar b$ invariant mass). Concerning the viable Yukawa structures of the N2HDM, we have found that all such anomalies (with the $\tau^+\tau^-$ actually being rather marginal in the $\chi^2$ fits) can be explained simultaneously only through the $h_3$ mediated process in the case of Type-II and Type-Y. All such results have been obtained in the presence of all available theoretical and experimental constraints, which we have implemented through state-of-the-art numerical tools. Furthermore, we have also found that the regions of the N2HDM parameter space explaining the discussed anomalous data are not excessively fine-tuned, in comparison to what is obtained in more minimal models of the Higgs sector. Therefore, the N2HDM tested here is of phenomenological interest, so that we have finally produced several BPs, covering both the Type-II and Type-Y Yukawa structure, amenable to further investigation. From an experimental perspective, the most direct tests of this interpretation are improved searches in the $\gamma\gamma b\bar b$ and $\tau^+\tau^- b\bar b$ channels, together with dedicated analyses of the inverted $b\bar b\,\gamma\gamma$ and $\gamma\gamma\tau^+\tau^-$ final states, whose rates are tightly correlated in the viable parameter region and therefore provide a concrete avenue to validate or falsify the scenario at Run~3 and the HL-LHC. Ultimately, our findings motivate continued theoretical and experimental efforts to explore the rich phenomenology of the N2HDM and its potential to shed light on BSM physics. 
		
		\section*{Acknowledgments}
		\sloppy
		MB acknowledges the support of Narodowe Centrum Nauki under OPUS grant no. 2023/49/B/ST2/03862. SM is supported in part through the NExT Institute and STFC CG ST/X000583/1.

\bibliography{main} 
\bibliographystyle{JHEP}
	\end{document}